\documentclass[prd,nofootinbib]{revtex4}
\usepackage{epsfig,amsmath}

\def\Qi{{\bf Q}_i}

\def\Qz{{\bf Q}}
\def\Q1{{\bf Q}_1}
\def\QN{{\bf Q}_N}

\def\qj{{\bf q}_j}
\def\phiv{\mbox{\boldmath$\phi$}}
\def\lambdav{\mbox{\boldmath$\lambda$}}
\def\phivs{\mbox{\boldmath${\scriptstyle{\phi}}$}}
\def\muv{\mbox{\boldmath$\mu$}}
\def\muvs{\mbox{\boldmath${\scriptstyle{\mu}}$}}
\def\d{{\rm d}}

\def\e{{\rm e}}
\def\i{{\rm i}}
\def\gs{\gamma_S}
\def\gss{\gamma_s}
\def\gq{\gamma_q}
\def\ls{\lambda_S}
\def\vexp{$VT^3 \, \exp[-0.7 \, {\rm GeV}/T]$}

\def\ss{${\rm s}\bar{\rm s}\;$}
\def\ssb{\langle {\rm s}\bar{\rm s}\rangle}
\def\uub{\langle {\rm u}\bar{\rm u}\rangle}
\def\ddb{\langle {\rm d}\bar{\rm d}\rangle}
\def\ppb{${\rm p}\bar{\rm p}\;$}

\def\kpi{$\langle {\rm K}^+ \rangle / \langle \pi^+ \rangle$}
\def\piN{$\langle \pi^- \rangle / \langle {\rm N}_p \rangle$}

\begin{document}

\title{Chemical equilibrium study in nucleus-nucleus collisions 
at relativistic energies} 

\author{F. Becattini}\affiliation{Universit\`a di Firenze}
\affiliation{INFN Sezione di Firenze, Florence, Italy} 
\author{M. Ga\'zdzicki}\affiliation{Institut f\"ur  Kernphysik, Universit\"at 
Frankfurt, Frankfurt, Germany}\affiliation
{\'Swi\c{e}tokrzyska Academy, Kielce, Poland}
\author{A. Ker\"anen}\affiliation{University of Oulu, Oulu, Finland}
\affiliation{INFN Sezione di Firenze, Florence, Italy}
\author{J. Manninen}\affiliation{University of Oulu, Oulu, Finland}
\author{R. Stock}\affiliation{Institut f\"ur  Kernphysik, Universit\"at Frankfurt, 
Frankfurt, Germany}\affiliation{CERN, Geneva, Switzerland}

\begin{abstract}
We present a detailed study of chemical freeze-out in nucleus-nucleus collisions 
at beam energies of 11.6, 30, 40, 80 and 158$A$ GeV. By analyzing hadronic 
multiplicities within the statistical hadronization approach, we have studied
the strangeness production as a function of centre of mass energy and of
the parameters of the source. We have tested and compared different versions of 
the statistical model, with special emphasis on possible 
explanations of the observed strangeness hadronic phase space under-saturation. 
We show that, in this energy range, the use of hadron yields at midrapidity
instead of in full phase space artificially enhances strangeness production and 
could lead to incorrect conclusions as far as the occurrence of full chemical 
equilibrium is concerned. In addition to the basic model with an extra strange 
quark non-equilibrium parameter, we have tested three more schemes: a two-component 
model superimposing hadrons coming out of single nucleon-nucleon interactions 
to those emerging from large fireballs at equilibrium, a model with local 
strangeness neutrality and a model with strange and light quark non-equilibrium 
parameters. The behaviour of the source parameters as a function of colliding 
system and collision energy is studied. The description of strangeness 
production entails a non-monotonic energy dependence of strangeness saturation
parameter $\gs$ with a maximum around 30$A$ GeV. We also present predictions of 
the production rates of still unmeasured hadrons including the newly
discovered $\Theta^+(1540)$ pentaquark baryon.    
\end{abstract}

\maketitle

\section{Introduction}

The main goal of the ultra-relativistic nucleus-nucleus (A-A) collisions programme 
is to create in terrestrial laboratories a new state of matter, the Quark-Gluon 
Plasma (QGP). The existence of this phase, where quarks and gluons are deconfined, 
i.e. can freely move over several hadronic distances, is a definite prediction of  
quantum chromodynamics (QCD). In a search for QGP signals A-A collisions at 
different centre of mass energies per nucleon-nucleon (NN) pair have been studied: 
from few GeV to several hundreds of GeV recently attained in Au-Au collisions at RHIC. 

Recently, accurate measurements of hadron production in central Pb-Pb collisions 
at 40, 80 and 158$A$ GeV of beam energy became available \cite{Af2002mx} and 
also preliminary data at 30$A$ GeV have been presented \cite{Alt2003rn} 
following an energy scan programme carried out by the experiment NA49 at CERN SPS. 
This programme is motivated by the hypothesis \cite{Gaz1998vd} that the threshold for 
creation of QGP in the early stage of Pb-Pb collisions might be located in the low
SPS energy range, roughly between 20 and 40$A$ GeV of beam energy.

One of the main results of the study of high energy A-A collisions is a surprising 
success of the statistical-thermal models in reproducing essential features of 
particle production \cite{csatz,sollf,bgs,pbm,yen,beca01,pbmrhic,flork}. 
This model succeeds also in describing
particle multiplicities in many kinds of elementary collisions \cite{beca,beca2,becapt}, 
suggesting that statistical production is a general property of the hadronization 
process itself \cite{beca2,vari}. Furthermore, the statistical hadronization model (SHM) 
supplemented with the hydrodynamical expansion of the matter, to a large extent also 
reproduces transverse momentum spectra of different particle species \cite{marco}. 

Hence, the SHM model proves to be a useful tool for the analysis of soft 
hadron production and particularly to study strangeness production, whose enhancement 
has since long been proposed as a signature of QGP formation. Furthermore, anomalies 
in the energy dependence of strangeness production have been predicted as a signature of
deconfinement and have been indeed observed experimentally \cite{anomalies}, 
suggesting that the onset of the phase transition could be located around 30$A$ GeV.  
It is thus important to make a systematic analysis, within the framework of SHM,
of the presently available hadronic multiplicities measured in Pb-Pb collisions at 30, 
40 and 80$A$ GeV, which - to our knowledge - is done here for the first time.

Along with these intriguing questions, our work is also motivated by issues related
to the application of statistical model itself. In fact, different versions of this
model have been used in the past by different authors leading to somewhat different 
results and conclusions. 
These mainly stem from the alternative use of midrapidity and full phase 
space multiplicities, from the allowance of non-equilibrium abundances of hadrons, 
from the assumption of exact local vanishing strangeness etc. Therefore,
we consider the comparison of these different approaches a worthwhile step. This has been
made it possible by now by the availability of an accurate and large multiplicity 
sample in Pb-Pb collisions at 158$A$ GeV as well as the corresponding data for pp 
interactions.   

The paper is organized as follows: a brief description of our main version of the SHM 
is given in Sect.~2. The experimental data selected for the analysis are summarized 
in Sect.~3. In Sect.~4 the results of the analysis using the main version and alternative
schemes of the SHM are given. Finally, in Sect.~5 we present and discuss the energy 
dependence of the chemical freeze-out stage. Summary and conclusions are drawn in 
Sect.~6.

\section{The Statistical Hadronization Model}

The main idea of the SHM is that hadrons are emitted from regions at statistical 
equilibrium, called clusters or fireballs. No hypothesis is made about how 
statistical equilibrium is achieved; this can be a direct consequence of the 
hadronization process. In a single collision event, there might 
be several clusters with different collective momenta, different
overall charges and volumes. However, Lorentz-invariant quantities like particle 
multiplicities are independent of clusters momenta, while they depend on charges
and volumes. If final state interactions among formed hadrons occur, particle 
multiplicities are frozen when inelastic interactions cease (chemical freeze-out). 
Thus, by analyzing measured hadron abundances, a snapshot is taken of clusters 
at that particular stage of the evolution, which may significantly precede 
the final kinetic freeze-out stage, when also elastic interactions cease. 
However, it should be pointed out that chemical and kinetic freeze-out may
depend on the hadron species and the assumption of a single chemical freeze-out 
is certainly an approximation. Most calculations in SHM are carried out in
the framework of the ideal hadron-resonance gas, that is handling resonances 
as free particles: this amounts to take a considerable part of the 
hadronic interactions between strongly stable hadrons into account \cite{hage}.
  
As has been mentioned, final multiplicities depend on the distribution of 
initial conserved charges (baryon number, strangeness and electric charge) among 
the produced clusters. This distribution is determined by the dynamics of the
collision and is thus needed as an external input to the statistical model. However,
most analyses, including ours, are carried out by assuming a single fireball. 
This is possible provided that one of the two conditions below is fulfilled:
\begin{enumerate}
\item{} all clusters are large enough to allow a grand-canonical description 
and all of them have the same values of relevant intensive parameters, i.e. 
temperature and chemical potentials;
\item{} clusters are small and must be treated canonically (i.e. counting those
states having exactly the same charges as the cluster itself), yet they have the
same temperature and the distribution governing fluctuations of charges is the 
same as that obtained by splitting one large cluster - the {\em equivalent 
global cluster} EGC - having as volume the sum of all clusters rest frame volumes
and charges the sum of all clusters charges (see Appendix A). In this case the 
overall particle multiplicities turn out to be those calculated in the canonical,
perhaps grand-canonical, ensemble of the equivalent global cluster \cite{beca2}. 
The reduction to EGC could be achieved even for micro-canonical clusters with 
additional requirement on mass fluctuations \cite{becapt}.
\end{enumerate}
The first condition sets stronger requirements and applies in the Bjorken's 
boost-invariant scenario, where all clusters are to have the same parameters 
independently of their rapidity. The second condition is altogether weaker 
and leaves room for the compatibility between the single fireball analysis 
and a variation of net baryon number density in rapidity. This has been 
discussed in detail in ref.~\cite{bgs}. The argument can be summarized 
as follows: particle multiplicities, being Lorentz invariants, are unaffected 
by a shift in rapidity of the clusters; therefore, clusters arising from the 
splitting of the EGC can be ordered in rapidity according to their net baryon 
number without affecting fully integrated particle multiplicities and, at the same
time, giving rise to an effective variation of the baryon density profile.  
Although the second condition is certainly more appropriate in the examined energy 
range, from AGS to SPS, it must be pointed out that this should not be expected to 
precisely match physical reality, as well as the first condition in its domain of 
applicability. In other words, discrepancies (hopefully small) between calculations 
based on this model and measurement are to be expected, so that these analyses 
shall not provide perfect fits even though the statistical model was the underlying 
true model. 

In this paper we will stick to the picture outlined in the second condition,
which implicitely requires the use of full phase space multiplicities in order to 
(hopefully) integrating out correlations between clusters' momenta and charges. 
Besides their general fitness, $4\pi$ multiplicities also allow to safely enforce 
overall strangeness neutrality. As has been mentioned, if the second condition applies, 
the multiplicity of any hadron $j$ can be calculated in the canonical ensemble of 
the EGC. Hence, as the EGC has a much larger volume than single clusters', the 
grand-canonical ensemble, where charges are conserved on average, can be a good 
approximation (see Appendix A). This is the case for the collisions examined in 
this paper \cite{kerabeca}. In this case the mean {\em primary} multiplicity of the 
$j^{\rm th}$ hadron with mass $m_j$ and spin $J_j$ reads: 
\begin{equation}\label{mean}
\left< n_j \right> = \frac{(2J_j+1) V }{(2\pi)^3} \int \d^3 {\rm p} \; 
\left[ \e^{\sqrt{{\rm p}^2+m_j^2}/T-\muvs\cdot\qj/T} \pm 1 \right]^{-1}
\end{equation}
where $T$ is the temperature, $V$ the EGC volume, $\qj = (Q_j,B_j,S_j)$ is a 
vector having as components the electric charge, baryon number and strangeness 
of the hadron and $\muv = (\mu_Q,\mu_B,\mu_S)$ is a vector of the corresponding 
chemical potentials; the upper sign applies to fermions, the lower to bosons.
In order to correctly reproduce the data, it is also necessary to introduce at 
least one non-equilibrium parameter suppressing hadrons containing valence strange 
quarks, $\gamma_S \neq 1$ \cite{gammas}. With this supplementary parameter, hadron 
multiplicity is as in Eq.~(\ref{mean}) with the replacement:
\begin{equation}\label{fuga2}
\exp[\muv\cdot\qj/T] \rightarrow \exp[\muv\cdot\qj/T] \gamma_S^{n_s}
\end{equation}
where $n_s$ stands for the number of valence strange quarks {\em and} anti-quarks 
in the hadron $j$.  

The abundances of resonances is calculated convoluting (\ref{mean}) with a relativistic 
Breit-Wigner distribution over a mass interval $[m-\delta m, m+\delta m]$, 
where $\delta m = \min[m-m_{\mathrm{threshold}},\ 2\Gamma]$. The minimum mass 
$m_{\mathrm{threshold}}$ is needed to open all decay modes. Finally, the overall  
multiplicity to be compared with the data, is calculated as the sum of primary 
multiplicity (\ref{mean}) and the contribution from the decay of heavier hadrons:
\begin{equation}\label{branching}
\langle n_j \rangle  = \langle n_i \rangle^{\mathrm{primary}} + 
\sum_k \mathrm{Br}(k\rightarrow j) \langle n_k \rangle,
\end{equation}
where the branching ratios are taken from the latest issue of the Review of Particle 
Physics \cite{pdg} and the summation runs over decays which contribute to the 
experimentaly measured multiplicity. Among the hadrons and resonances contributing 
to the sum in Eq.~(\ref{branching}), in this work all known states quoted in 
ref.~\cite{pdg} up to a mass of 1.8 GeV are included (see discussion in Sect.~4).  

What we have hitherto described is the main version of the SHM used for the 
data analysis, that will be henceforth referred to as SHM($\gs$). As 
has been mentioned in the Introduction, in this work we also test other 
schemes and versions of the SHM, which will be described in detail in 
Sect.~4.

\section{Experimental Data Set}

The bulk of the experimental data consists of measurements made by NA49 
collaboration in central Pb-Pb collisions at beam momenta of 30, 40, 80 and 
158$A$ GeV, corresponding to $\sqrt s_{NN} =$ 7.6, 8.8, 12.3 and 17.2 GeV 
respectively.
The acceptance region in rapidity and transverse momentum covers a typical range 
from midrapidity to projectile rapidity and from 0 to 1.5 GeV/$c$ respectively.
The overall hadron multiplicities, quoted in referenced papers, were obtained 
using forward-backward symmetry in rapidity and by extrapolating the yields 
to full phase space. All results were corrected for the feed-down from weak 
decays, e.g. $\pi^-$ multiplicity does not include pions produced in decays of 
$\Lambda$ hyperons and K$^0_S$ mesons.

Central collisions were selected by a trigger using information from a downstream 
calorimeter, which measured the energy of the projectile spectator nucleons.  
Whilst at 30, 40 and 80$A$ GeV all published results refer to the 7.2\% most
central collision sample, at 158$A$ GeV different centrality selections (5\%, 
10\% and 20\% most central collisions) were used to measure various hadronic species. 
In this analysis we have rescaled all published multiplicities at 158$A$ GeV 
to the corresponding ones at 5\% most central collisions assuming that for the 
considered central collisions the hadron yield is proportional to the mean 
number of participant nucleona. The resulting scaling factors are 1.08 and 1.32 for 
10\% and 20\% most central collisions respectively \cite{Af2002mx,Af2002fk}. 

As far as AGS data at 11.6$A$ GeV is concerned, we have used both multiplicities 
measured by the experiments and extrapolations of measured rapidity distributions 
made in ref.~\cite{beca01} at 3\% top centrality. For $\Lambda$ we have made a 
weighted average of the multiplicities measured both at 5\% top centrality by 
E896 \cite{lamb896} and E891 \cite{lamb891}. For the former, the quoted 
experimental error was only statistical so that we have added a 10\% systematic 
error resulting in a value of 16.7$\pm$0.5$\pm$1.7. For the latter, we have 
used the extrapolated value in ref.~\cite{beca01} of 20.34$\pm$2.74. The error 
on the weighted average has been rescaled by 1.25 (i.e. $\sqrt{\chi^2}$) according 
to the PDG weighting method in case of discrepancy between different measurements 
\cite{pdg}. The obtained average has been rescaled by a factor 1.02 to convert 
it from 5\% to 3\% top centrality by assuming a linear dependence on the number 
of participants and by using the tables in ref.~\cite{protags}. Since the 
$\bar{\Lambda}$ to $\Lambda$ ratio has been measured only at midrapidity 
\cite{alamags}, we have obtained a $\bar\Lambda$ 4$\pi$ multiplicity assuming that 
the double ratio $(\langle \bar{\Lambda}\rangle/\langle\Lambda\rangle)_{y=0}/(\langle 
\bar{\Lambda} \rangle/\langle \Lambda \rangle)$ is the same at SPS and AGS 
energies. The final experimental  multiplicities and ratios used in our analysis are 
shown in tables~\ref{ags}, \ref{pbpb30}, \ref{pbpb40}, \ref{pbpb80} and \ref{pbpb158}.

In order to test the effect of the cut in rapidity on the resulting statistical 
model parameters (discussed in detail in Sect.~4) we have also determined the 
yields integrated over limited ($\Delta y = 1$ and $\Delta y =2$) rapidity windows 
around midrapidity in Pb-Pb collisions at 158$A$ GeV. This has been done by fitting 
the rapidity distributions measured by NA49 to a Gaussian or the sum of two Gaussians, 
with area and width as a free parameters and central values set to zero. The 
results are shown in table~\ref{yfit}. The quality of the fits is quite good, 
except for pions due to a couple of points near midrapidity; yet, this discrepancy 
does not affect significantly the integrated yield. In fact, it must be stressed 
that the main goal of these fits is to estimate an integral and not to reproduce 
accurately the shape of the distributions over the full measured range. We have also
checked that the extrapolations to full phase space are in good agreement with 
published measurements. 

\section{Analysis Results}

The analysis has been carried out by looking for the minima of the $\chi^2$:
\begin{equation}
    \chi^2 = \sum_i \frac{(n_i^{\rm exp} - n_i^{\rm theo})^2}{\sigma_i^2}
\end{equation}
where $n_i$ is the multiplicity of the $i^{\rm th}$ measured hadronic species and 
$\sigma_i$ is the sum in quadrature of statistical and systematic experimental 
error. 

The theoretical multiplicities are calculated according to Eq. (\ref{branching})
with the decay chain stopped to match the experimental definition of multiplicity 
to properly compare theoretical and experimental values. This occurs in Pb-Pb 
collisions after electromagnetic and strong decays and before weak decays, 
whilst in Au-Au collisions at AGS the weak decays of $\Lambda$, $\Sigma$, $\Xi$, 
$\Omega$ and K$^0_S$ are included.

The effect of the uncertainties on masses, widths and branching ratios of the 
involved hadrons on the fit parameters has been studied by the method described 
in ref.~\cite{becapt} and found to be negligible throughout.    

In order to cross-check our results and verify their robustness, we have performed 
the analysis with two independent numerical programs, henceforth referred to 
as A and B, which mainly differ with regard to the included resonances, their
decay modes and branching ratios.

\begin{figure}
\caption{Above: measured versus fitted multiplicities in the statistical model 
supplemented with $\gs$ parameter (SHM($\gs$)) in Au-Au collisions at a beam 
energy of 11.6$A$ GeV in the fit A; also quoted the best-fit parameters. Below: 
residual distribution. 
\label{agsf}}
\includegraphics[scale=0.9]{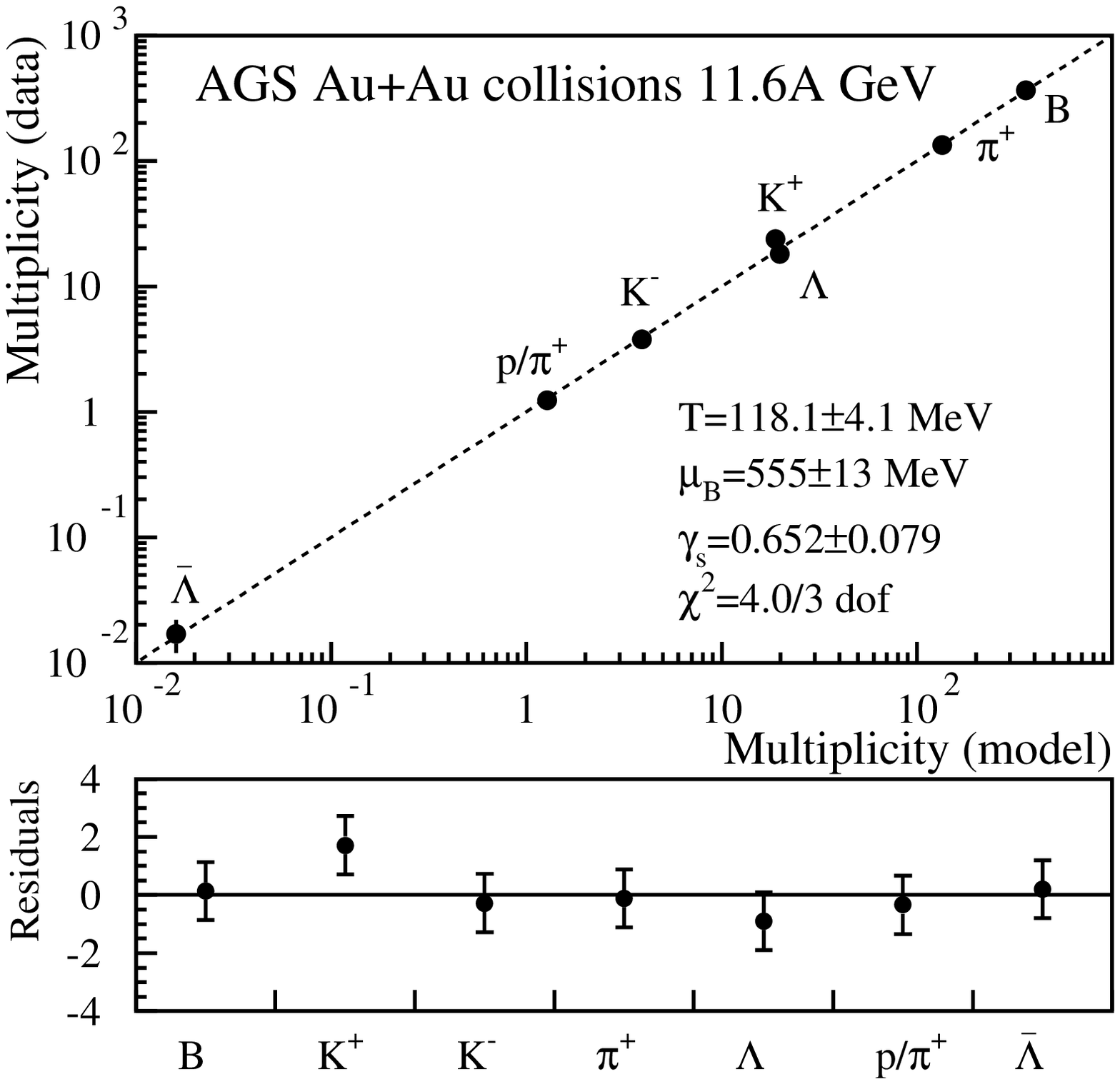}
\end{figure}

\begin{figure}
\caption{Above: measured versus fitted multiplicities in the statistical model 
supplemented with $\gs$ parameter in Pb-Pb collisions at a beam energy of 40$A$
GeV in the fit A; also quoted the best-fit parameters. Below: residual distribution. 
\label{pbpb40f}}
\includegraphics[scale=0.9]{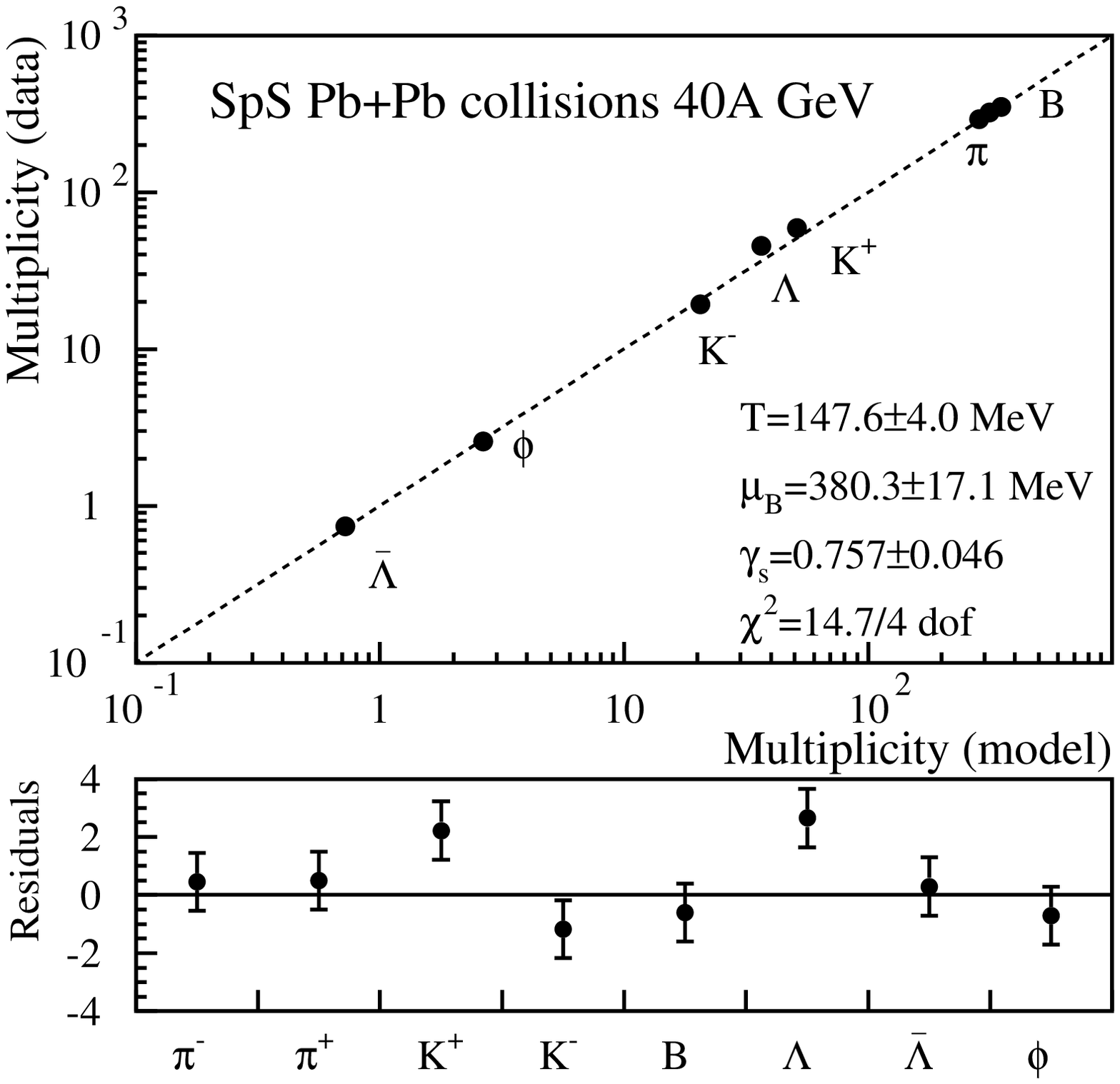}
\end{figure}

\begin{figure}
\caption{Above: measured versus fitted multiplicities in the statistical model 
supplemented with $\gs$ parameter in Pb-Pb collisions at a beam energy of 80$A$
GeV in the fit A; also quoted the best-fit parameters. Below: residual distribution. 
\label{pbpb80f}}
\includegraphics[scale=0.9]{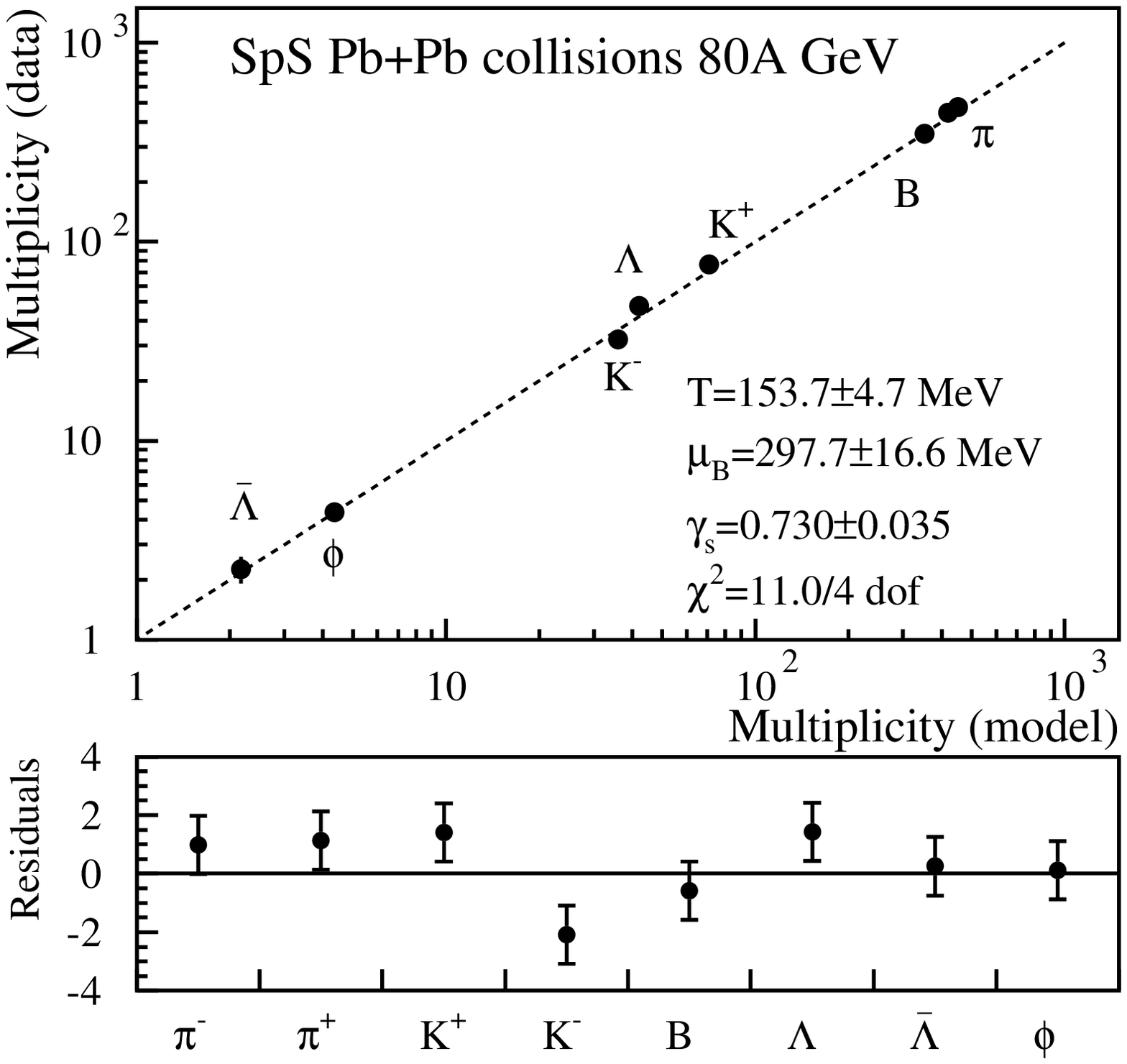}
\end{figure}

\begin{figure}
\caption{Above: measured versus fitted multiplicities in the statistical model 
supplemented with $\gs$ parameter in Pb-Pb collisions at a beam energy of 158$A$
GeV in the fit A; also quoted the best-fit parameters. Below: residual distribution.
Note that the $\Lambda(1520)$ was not used in the fit (see text). 
\label{pbpb158f}}
\includegraphics[scale=0.9]{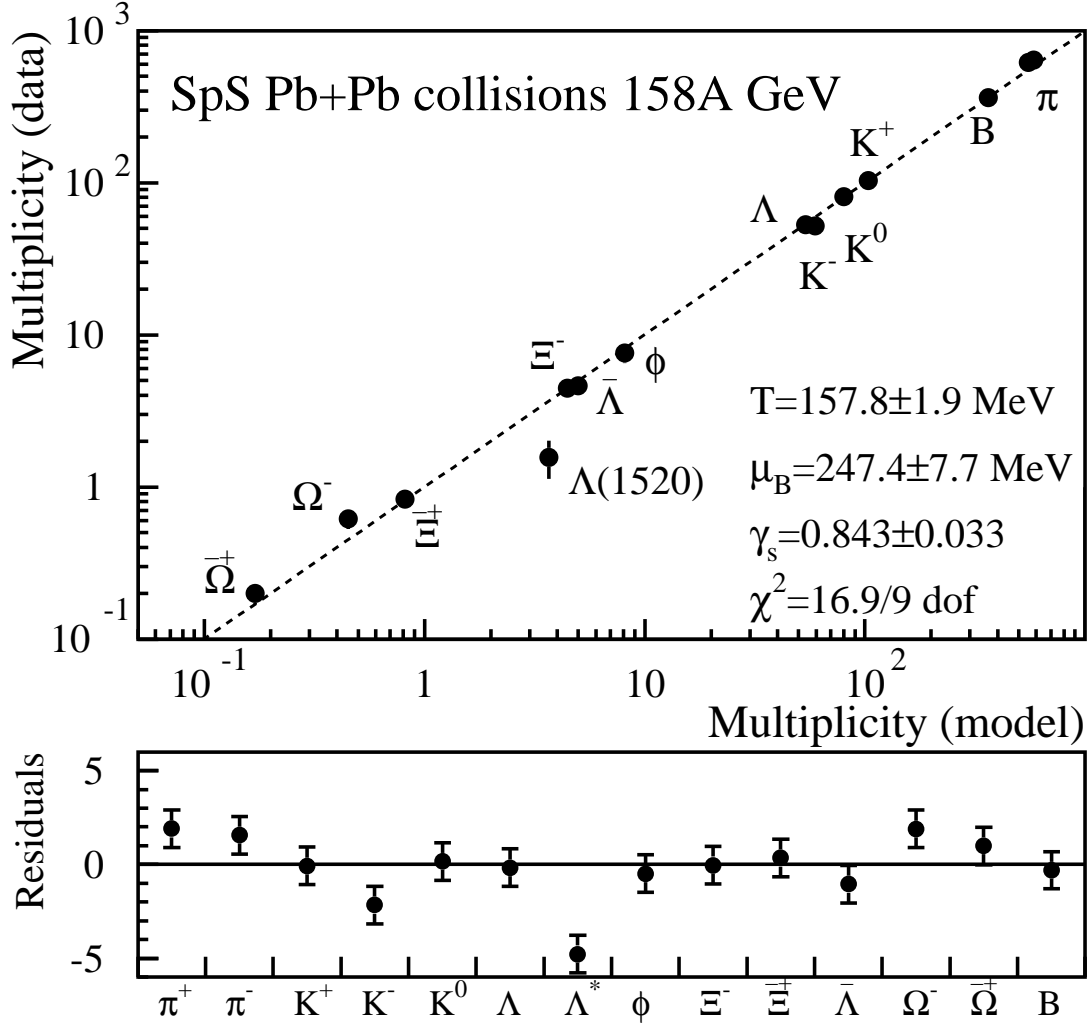}
\end{figure}

\begin{figure}
\caption{Comparison between measured and calculated (in fits A and B) \kpi and \piN 
ratios as a function of the centre-of-mass energy in the examined collisions. 
For the SPS energy points the statistical errors are indicated with solid lines, 
while the contribution of the common systematic error is shown as a dotted line
The lines connect the fitted values. \label{sqrts}}
\includegraphics[scale=1.6]{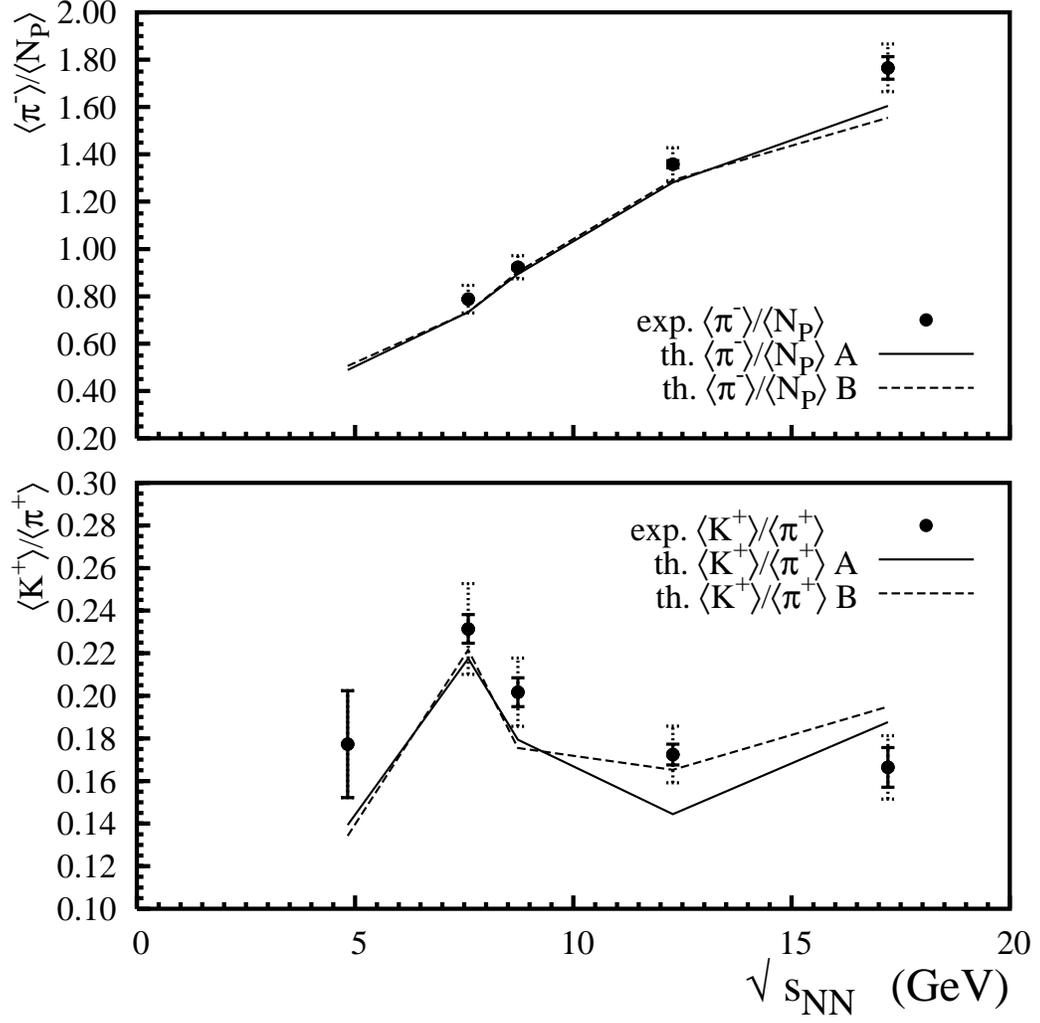}
\end{figure}

The fitted parameters within the main scheme SHM($\gs$) are shown in table~ 
\ref{parameters}, while the experimental and fitted multiplicities, along with the 
predicted yields of several hadron species are shown in tables~\ref{ags},
\ref{pbpb30}, \ref{pbpb40}, \ref{pbpb80} and \ref{pbpb158} and figs.~\ref{agsf},
\ref{pbpb40f}, \ref{pbpb80f} and \ref{pbpb158f}. We do not show any plot for
the fit in Pb-Pb collisions at 30$A$ GeV because of the low number of data
points. The quality of the fit is good 
throughout, as proved by the $\chi^2$'s values and we do not see any clear 
discrepancy between data and model, with the remarkable exception of the 
$\Lambda(1520)$ in Pb-Pb collisions at 158$A$ GeV. Due the 5$\sigma$ deviation
from the statistical model prediction, the measured $\Lambda(1520)$ yield has 
been removed from the fitted data sample as it could have biased the fit itself. 
We argue that this disagreement owes to its short lifetime ($\Gamma = 15.6$ MeV)
compared with all other used particles. If the kinetic freeze-out occurs after
some suitable delay from the chemical freeze-out, one can indeed justify the
low measured $\Lambda(1520)$ yield as the effect of the elastic reinteractions  
of its decay products. The quality of the fits is further illustrated in
fig.~\ref{sqrts} where the measured and fitted ratios \piN and \kpi are plotted 
as a function of $\sqrt{s_{NN}}$; these ratios have been chosen as it has been 
proposed that their energy depedence plays an important role in the search 
for deconfinement onset at SPS energies \cite{Gaz1998vd}.

The observed differences in the fit parameters between A and B are of the order 
of the fit errors. They may be considered as an estimate of the systematic error 
due to uncertainties in the implementation of the model. The first set of parameters 
in table~\ref{parameters} have been obtained by using the full data, whilst the 
second set has been obtained by using the maximal common set of particles measured 
in the four collisions, that is $\pi^+$, K$^+$, K$^-$, $\Lambda$, $\bar\Lambda$ 
and the participant nucleons (net baryon number) $B$. By comparing fit results 
in the same analysis (A with A and B with B), it can be seen that the obtained
parameters are in good agreement and only in one case a discrepancy larger than
one standard deviation ($\gs$ in Pb-Pb at 158$A$ GeV) is observed; this demonstrates 
the robustness of the results. 

We have also included in tables~\ref{ags}, \ref{pbpb30}, \ref{pbpb40}, \ref{pbpb80} 
and \ref{pbpb158} the prediction for the yield of the recently discovered 
$\Theta^+$ pentaquark baryon (uudd$\bar{\rm s}$) by using as input mass $m=1540$ 
MeV and $J=1/2$. According to the SHM($\gs$) model, in the Boltzmann limit this 
simply reads:
\begin{equation}\label{theta}
\left< n_{\Theta^+} \right> = \frac{\gs V}{\pi^2} m^2 T 
{\rm K}_2 \left(\frac{m}{T}\right) \exp[\mu_B/T+\mu_Q/T+\mu_S/T]  
\end{equation}
if we disregard feeding from possible excited states.

As the number of data points in Pb-Pb collisions at 30$A$ GeV was not sufficient 
to determine the four free parameters unambiguously, we have forced $T$ to lie 
on the parabolic chemical freeze-out curve (\ref{tmuparam}) in Sect.~5, 
interpolating the other four points in the $\mu_B-T$ plane. This method has proved 
to be able to provide unambiguous solutions for the remaining three free parameters.

The first quoted error beside the best-fit value in table~\ref{parameters} is the 
error coming out from the fitting program (inferred from the analysis of the 
$\chi^2=\chi^2_{\rm min} +1$ level contours) whereas the second error is the fit 
error rescaled by a factor $\sqrt{\chi^2_{\rm min}/dof}$ where $dof$ is the 
number of degrees of freedom. We deem that the latter is a more realistic 
uncertainty on the parameters with respect to the fit error because of the 
"imperfect" $\chi^2_{\rm min}/dof$ values, expected to be 1 on average
if the model correctly matched physical reality. The argument, which is the same
used in the Particle Data Book \cite{pdg} when averaging discrepant data, is 
as follows: if $\chi^2_{\rm min}/dof \neq 1$, then the model cannot reproduce 
the data at the level of accuracy relevant to the experimental errors; on the 
other hand, this would be the case if experimental errors were larger and, 
particularly, if they were rescaled by a common factor $S$ so that:
\begin{equation}
    \chi^{2'} = \sum_i \frac{(n_i^{\rm exp} - n_i^{\rm theo})^2}{(S\sigma_i)^2}=
    \frac{\chi^2}{S^2}
\end{equation}
With this simple rescaling of the $\chi^2$, the best-fit parameters would be
unchanged, whereas their relevant errors would scale up by a factor $S$. In 
fact, the new covariance matrix $C'$ for the parameter vector $X$ is related to 
the $\chi^{2'}$ around the minimum through:
\begin{equation}
  \chi^{2'}(X) = \chi^{2'}_{\rm min} + (X-X_0)^T {\sf C'}^{-1}(X-X_0) 
\end{equation}
Also:
\begin{equation}
  \chi^2(X) = \chi^2_{\rm min} + (X-X_0)^T {\sf C}^{-1} (X-X_0) 
\end{equation}
and, being $\chi^{2'}=\chi^2/S^2$, one finds ${\sf C}' = S^2 {\sf C}$.
Therefore, since this covariance matrix would be the outcome of a standard 
quality fit, with $\chi^{2'}_{\rm min}/dof = 1$, the errors rescaled by $S$ 
can be regarded as the sensible minimal uncertainty on the parameters.  
These rescaled errors have indeed been used in all of the plots in this paper. 

\begin{figure}
\caption{Spectrum of known light-flavoured hadronic species up to a mass of 1.8 GeV.
\label{spectr}}
\includegraphics[scale=0.9]{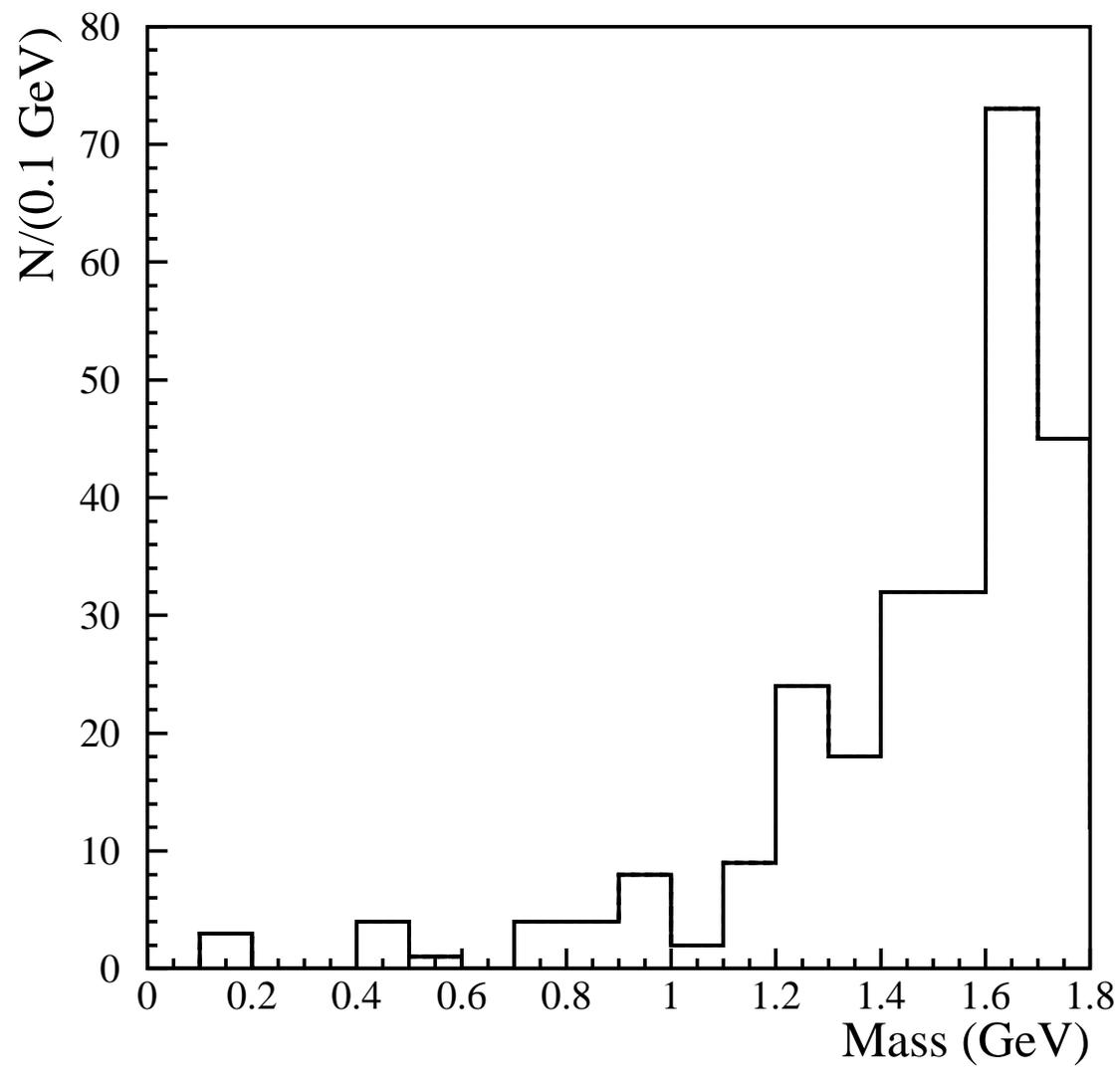}
\end{figure}

\begin{figure}
\caption{Left: primary yields of various particles as a function of the cut-off
on the hadronic mass spectrum. Right: fitted $\gs$, baryon-chemical potential and
temperature as a function of the cut-off on the hadronic mass spectrum. \label{stabil}}
\includegraphics[scale=1.6]{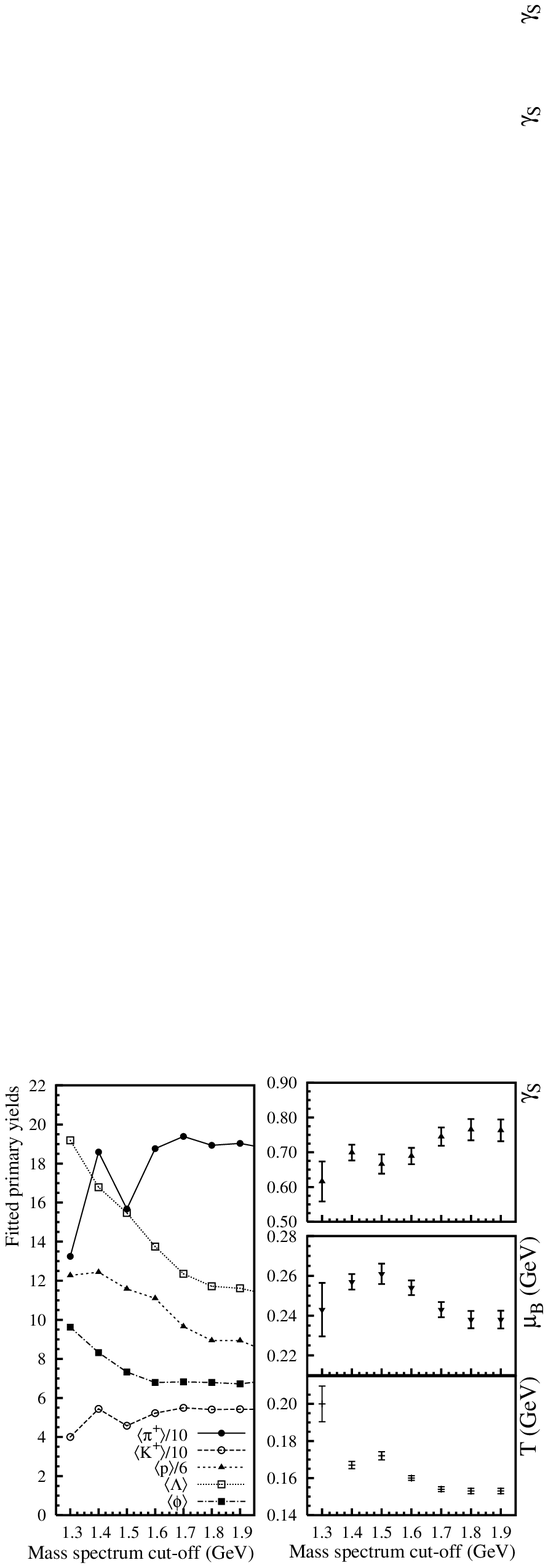}
\end{figure}

A major issue in the multiplicity fits is where to stop the inclusion of heavy
light-flavoured resonances contributing to measured particle yields in
Eq.~(\ref{branching}). The relevance of this cut-off is owing to the peculiar 
shape of the hadron mass spectrum, which rises almost exponentially between 1 
and 1.7 GeV and drops thereafter probably due to the missing experimental 
information (see fig.~\ref{spectr}). Should the 
number of states keep on increasing exponentially, the problem is set of the 
physical meaning of the obtained parameters, which could be heavily affected 
by the ignorance of further hadronic states. In fact, although the production of 
resonances decreases exponentially with the mass, the effect on secondary light 
particles through the decay chain could be balanced and even exceeded by the 
increasing number of states. We have thus checked the stability of the obtained 
parameters in the four collisions by varying the cut-off on the mass spectrum 
in a range where we are reasonably confident on the complete experimental 
knowledge and the number of states apparently rises exponentially (i.e. up to
1.7-1.8 GeV) and repeating the fit. As shown in fig.~\ref{stabil}, the fitted
$T$, $\mu_B$ and $\gs$ in Pb-Pb at 158$A$ GeV are indeed fairly constant from 1.3 to 
1.9 GeV. Furthermore, the outcoming primary yields of some measured particles 
tend to saturate at cut-off masses of about 1.8 GeV, implying that the 
contribution of resonance decays to secondaries (needed to keep the final 
multiplicity close to the measured value) settles down as well and 
the inclusion of heavier states yields a more and more negligible contribution. 
This is a clear indication of the significance of the fit results. 
A similar pattern occurs in all other examined collisions. 

A major result of these fits is that $\gs$ is significantly smaller than 1
in almost all cases (with a possible exception at 30$A$ GeV, see table~
\ref{parameters}), that is strangeness seems to be under-saturated with respect 
to a completely chemically equilibrated hadron gas. This confirms previous 
findings \cite{bgs,beca01,cley}. There is a 
considerable interest and ongoing investigations about this deviation of 
the data from the fully equilibrated hadron gas, particularly motivated
by the fact that strangeness production is considered as a possible QGP 
signature. It is therefore worth to examine and test, with the presently 
available large data sample, different scenarios which have been put forward 
to account for the under-saturation of strangeness.

\subsection{Full equilibrium and midrapidity ratios}

\begin{figure}
\caption{Rapidity distributions of $\pi^{\pm}$, K$^\pm$, $\phi$ and $\Omega$ emitted
from a single fireball at rest at a kinetic freeze-out temperature of $T=125$ MeV. 
\label{rapid}}
\includegraphics[scale=0.9]{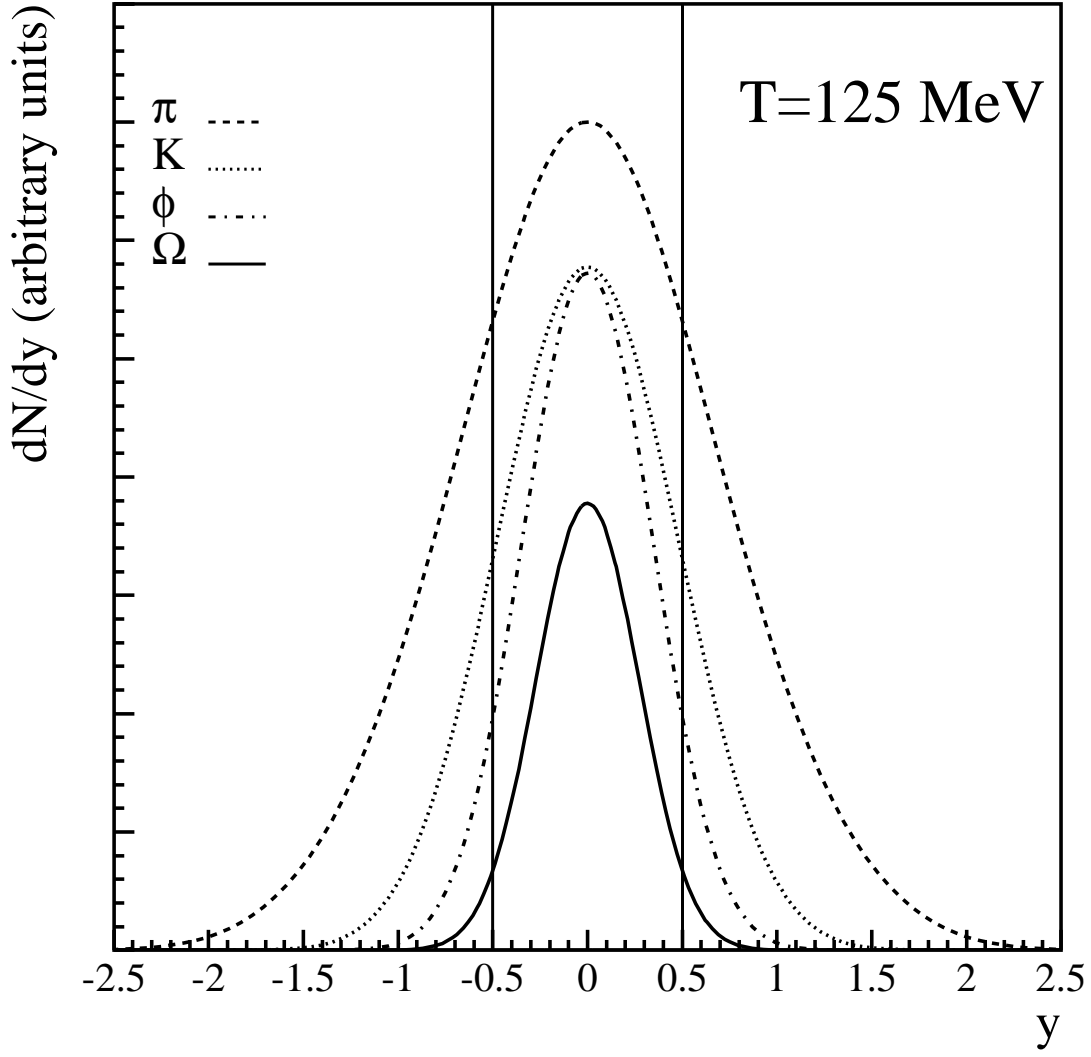}
\end{figure}

As we have seen, fits to full phase space multiplicities within the SHM 
yield $\gs < 1$ in most cases. However, good tests of the same model without 
extra strangeness suppression (i.e. assuming $\gs = 1$) have been obtained by 
fitting ratios of hadronic yields within a limited rapidity range around 
midrapidity at top SPS energy \cite{pbm}. 
This is an appropriate method of estimating the parameters of the sources only 
if the boost-invariant Bjorken scenario holds, at least as a good approximation, 
over a large rapidity interval (say $\Delta y \simeq 6$) because, in this case, 
particle ratios at midrapidity are the same as in full phase space. However, 
rapidity distributions of hadrons at SPS energies do not feature boost-invariance 
\cite{Af2002mx,Af2002fk,Afuu,Af2002ub,Afuu,Af2002he} and a cut at midrapidity can 
artificially enhance heavy particle yield with respect to light ones (see fig.~ 
\ref{rapid}), as long as their kinetic freeze-out occurs at the same temperature 
and the leading baryon effect can be neglected.
In the statistical model of a single fireball this can be easily understood, for 
the width of the rapidity distribution decreases as a function of mass according to 
(in the Boltzmann approximation):
\begin{equation}
   \frac{\d N}{\d y} \propto \left( m^2 T + \frac{2 m T^2}{\cosh y} + \frac{2 T^3}
    {\cosh^2 y} \right) \exp[-m \cosh y/T]
\end{equation} 
Yet, it is worth testing the effect of the rapidity cut on measured distributions 
rather than using arguments based solely on the statistical model. Therefore, we 
have fitted, within the scheme A, the integrated yields over limited rapidity windows 
measured in central Pb+Pb collisions at 158$A$ GeV by NA49 (see Sect.~3 for details) 
and quoted in table~\ref{deltay2} as well as the yields measured by WA97 \cite{wa97}
over a $\Delta y = 1$ window around midrapidity. 
We first note that, according to table~\ref{deltay2}, the integrated yields over 
$\Delta y = 1$ measured by NA49 and WA97 are in good agreement with each other. 
Since the fit to the statistical model gave $\gs \simeq 1$ \cite{becasqm} for WA97 
data, the same is expected for the integrated NA49 yields over the same rapidity 
window. This is indeed what we find, as shown in table~\ref{deltay}. While temperature 
and baryon-chemical potential are essentially unchanged, the best-fit value of $\gs$ is
closer to 1 than that obtained in full phase space in fit A 
(see table~(\ref{parameters}) and it is also compatible with 1 within the error. 

We then conclude that the superfluity of $\gs$ in analysis of midrapidity 
particle yields, at least at top SPS energy, is likely to owe to the artificial 
enhancement of strange particles with respect to lighter non-strange ones,
induced by the cut on rapidity. The fact that $\gs \simeq 1$ for midrapidity
yields is then not an indication of a fully equilibrated hadron gas at 
midrapidity; even if such equilibrated fireball existed at the estimated 
kinetic freeze-out temperature of $T \approx 125$ MeV \cite{kfo}, the $\Delta y = 1$ 
window would be too narrow for a correct estimation of chemical freeze-out
parameters (see fig.~\ref{rapid}) because lighter particles would be cut down
significantly. 

\subsection{Strangeness correlation volume}

To account for the observed under-saturation of strangeness, a picture has been
put forward in which strangeness is supposed to be exactly vanishing over 
distances less than those implied by the overall volume $V$ \cite{redl}. 
We henceforth refer to this version of the statistical model as SHM(SCV). 
Following the description of the model in Sect.~2, this means that the produced 
clusters or fireballs emerge with $S=0$ and they are not allowed to share non-vanishing 
net strangeness. Assuming, for sake of simplicity, that all clusters have the same 
typical volume $V_c$ and that the equivalence of the set of clusters to a global 
fireball still applies for baryon number and electric charge (but not to strangeness) 
the following expression of the primary average multiplicities can be obtained 
(see Appendix B):
\begin{equation}\label{scanonical}
 \langle n_j \rangle = \frac{V}{V_c} \frac{(2J_j+1)V_c}{(2\pi)^3} 
 \sum_{n=1}^\infty \int \d^3 {\rm p} \;(\mp 1)^{n+1} 
 \exp[-n\sqrt{{\rm p}^2 + m_j^2}/T + n \mu_B B_j/T + n \mu_Q Q_j/T] 
 \frac{Z_c(-nS_j)}{Z_c(0)} 
\end{equation}
where
\begin{equation}
 Z_c = \frac{1}{2\pi} \int^\pi_{-\pi} \d \phi \; \exp \Big[ \sum _j 
 \frac{(2J_j+1)V_c}{(2\pi)^3} \int \d^3 {\rm p} \; 
 \log (1 \pm \e^{-\sqrt{{\rm p}^2 + m_j^2}/T +  \mu_B B_j/T + \mu_Q Q_j/T 
 -\i \phi S_j})^{\pm 1} \Big]
\end{equation}
is the so-called {\em strange canonical partition function} of a single cluster.
As usual, in the above equations, the upper sign is for fermions and the lower
for bosons.

If $V_c$ is sufficiently small, the multiplicities of strange hadrons turn out to
be significantly suppressed with respect to the corresponding grand-canonical
ones due to the enforcement of exact strangeness conservation in a finite system, 
an effect called {\em canonical suppression}. Furthermore, the suppression
features hierarchy in strangeness, namely it is stronger for $\Omega$ ($S=3$) and
$\Xi$ ($S=2$) than for kaons and $\Lambda$'s, so it can be argued that this 
can account for the actually observed hierarchical pattern of extra strangeness
suppression which goes like $\gs^{|S|}$ for open strange particles. The 
discriminating difference between this picture and our main scheme SHM($\gs$) 
described in Sect.~2 is concerned with hidden strange particles such as $\phi$, 
which do not suffer canonical suppression, so that its theoretical multiplicity 
in SHM(SCV) turns out to be simply the same as in a cluster with volume $V$, 
that is given by the formula (\ref{mean}) without $\gs^2$ suppression.   

We have made a test of this model by fitting the data sample of full phase
space multiplicities in Pb-Pb collisions at 158$A$ GeV fixing $\gs=1$ and 
determining the parameters $T$, $V$, $\mu_B$ and $f = V_c/V$ within the scheme A. 
The results are shown in tables~\ref{models} and \ref{pbpb158}. The quality of the fit
is worse with respect to the SHM($\gs$) model mainly because of the underestimated
pion yield and the larger of $\phi$. The latter is expected, as has been mentioned. 
As far as pion discrepancy is concerned, the deviation stems
from the very fact that they are the only non-strange particles in the fit. The
minimization procedure tries to accommodate the relative ratios among strange
hadrons by fixing $V_c$ and $T$, then it tries to set the overall normalization
$V$ and at this stage a competition sets in between the set of strange and 
non-strange particles. Since pions are only two entries, the fit prefers 
to keep them low rather than raising the whole set of strange particles.       

Our result suggests that, for the local strangeness correlation to be an 
effective mechanism, the cluster volume should be of the order of 2.5\% of the 
overall volume. Otherwise stated, strange quarks should have not propagated beyond 
a distance of about 4 fm from the production point up to chemical freeze-out, 
if we take the overall volume of about 3 10$^3$ fm$^3$ as coming out from this 
fit where hadrons are pointlike particles.      

\subsection{Superposition of NN collisions with a fully equilibrated fireball}

In this picture, henceforth referred to as SHM(TC), the observed hadron production 
is approximately the superposition of two components (TC): one originated from one large 
fireball at complete chemical equilibrium at freeze-out, with $\gs=1$, and another 
component from single nucleon-nucleon collisions. In fact, according to simulations 
based on transport models, a significant fraction of beam nucleons interacts only 
once with target nucleons \cite{urqmd}. With the simplifying assumption of disregarding 
subsequent inelastic collisions of particles produced in those primary NN collisions, 
the overall hadron multiplicity can be written then as:
\begin{equation}
 \langle n_j \rangle = \langle N_c \rangle \langle n_j \rangle_{NN} +
 \langle n_j \rangle_V
\end{equation}
where $\langle n_j \rangle_{NN}$ is the average multiplicity of the $j^{\rm th}$
hadron in a single NN collision, $\langle N_c \rangle$ is the mean number of
single NN collisions giving rise to non-re-interacting particles and $\langle n_j 
\rangle_V$ is the average multiplicity of hadrons emitted from the equilibrated
fireball, as in Eq.~(\ref{mean}), with $\gs=1$. The $\langle n_j \rangle_{NN}$ term 
can be written in turn as:
\begin{equation}
 \langle n_j \rangle_{NN} = \frac{Z^2}{A^2} \langle n_j \rangle_{pp} +
 \frac{(A-Z)^2}{A^2} \langle n_j \rangle_{nn} + \frac{2Z(A-Z)}{A^2}
 \langle n_j \rangle_{np}
\end{equation}

Since it is known that in NN collisions strangeness is strongly suppressed \cite{beca2}
the idea is to ascribe the observed under-saturation of strangeness in heavy
ion collisions to the NN component, leaving the central fireball at complete
equilibrium, i.e. with $\gs=1$. Of course, this is possible provided that $\langle
N_c \rangle$ is sufficiently large. This production mechanism has probably some 
consequences on the final rapidity and momentum distributions of the different
species, whose calculation goes certainly beyond the scope of this paper. Instead,
we have confined ourselves to integrated multiplicities and tried to fit
$T$, $V$, $\mu_B$ of the central fireball and $\langle N_c \rangle$ by using 
NA49 data in Pb-Pb collisions at 158$A$ GeV within the scheme A. 

\begin{figure}
\caption{Above: measured versus fitted multiplicities in the statistical model 
supplemented with strangeness suppression in pp collisions at a beam energy of 
158 GeV corresponding to $\sqrt s = 17.2$ GeV; also quoted the best-fit parameters. 
Below: residual distribution. \label{pp160f}}
\includegraphics[scale=0.9]{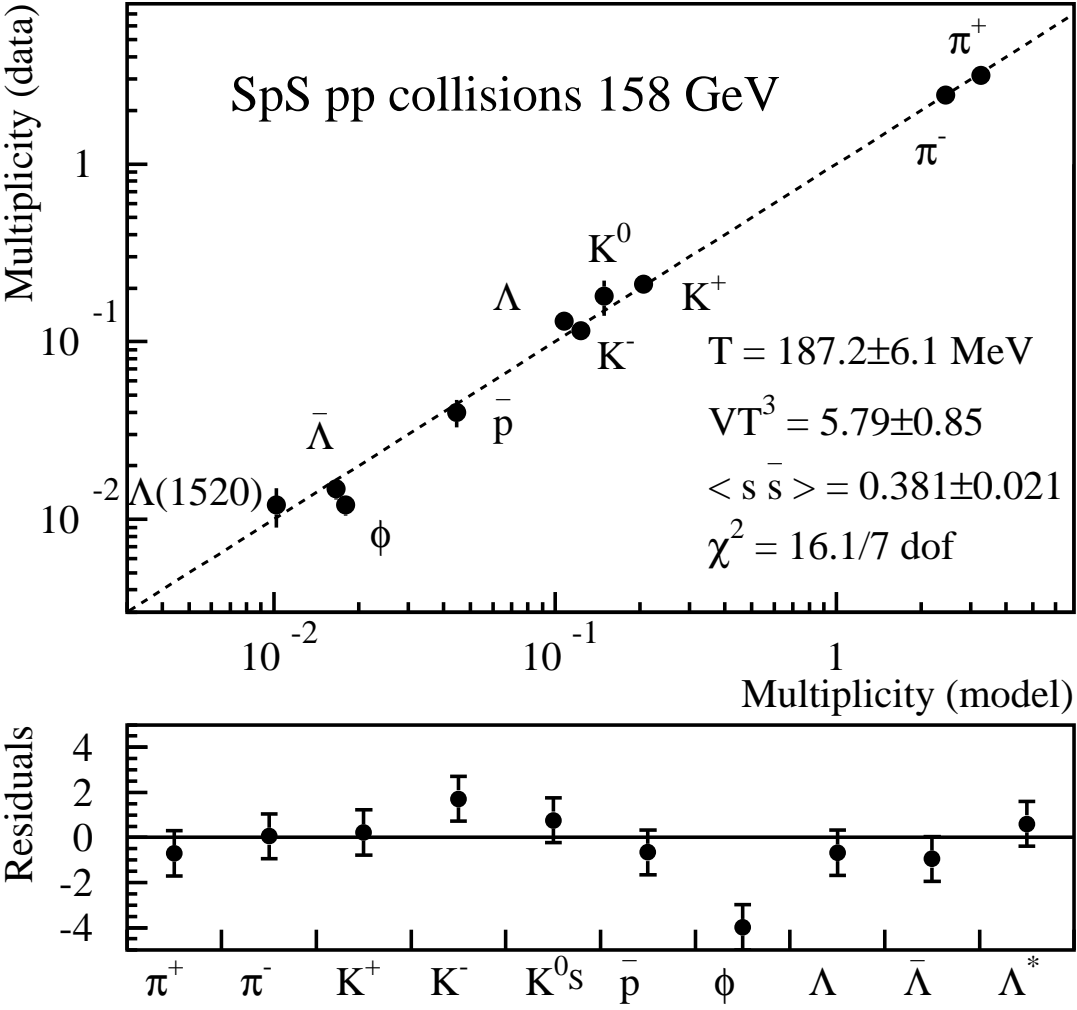}
\end{figure}

To calculate $\langle n_j \rangle_{NN}$ we have used the statistical model and
fitted pp full phase space multiplicities measured at $\sqrt s = 17.2$ GeV (i.e.
the same beam energy) by the same NA49 experiment. For np and nn collisions, the 
parameters of the statistical model determined in pp are retained and the 
initial quantum numbers are changed accordingly. Theoretical multiplicities have 
been calculated in the canonical ensemble, which is described in detail in 
ref.~\cite{becapt}. Instead of the usual 
$\gs$ parametrization, the new parametrization described in ref.~\cite{becapt} has 
been used in which one assumes that some number of \ss pairs, poissonianly 
distributed, hadronizes; the extra strangeness suppression parameter $\gs$ is thus 
replaced by the mean number of these \ss pairs, $\ssb$.\\ 
The results of the fit are shown in table~\ref{pp160} along with fitted and
predicted hadron multiplicities, including the $\Theta(1540)$ pentaquark baryon,
and in fig.~\ref{pp160f}. The temperature value is 
significantly higher than in pp and \ppb collisions at higher energy, an effect
already observed for center-of-mass energies below 20 GeV \cite{beca2,becapt}.
We conjecture that this is a possible indication of a beginning inadequacy of
the canonical ensemble at low energy, where exact conservation of energy and 
momentum should start to play a significant role. Perhaps this is the point
where the microcanonical hadronization of each cluster is a more appropriate
approach.

The results of the fit to Pb-Pb collisions are shown in table~\ref{models}. 
The fit quality, as well as the obtained values of $T$, $\mu_B$, are comparable 
to the main fit within the SHM($\gs$) model. The predicted number of
"single" NN collisions is about 50 with a 16\% uncertainty. Thus, only 260 
nucleons out of 360 contribute to the formation of large equilibrated fireballs.
The percentage of primary hadrons stemming from NN collisions varies from 14\%
for pions to 27\% of $\rho$'s and protons and to 0.5\% of $\Omega$'s.       
It should be pointed out that the fitted parameters are affected by a
further systematic error owing to the uncertainty on the parameters of the
statistical model in NN collisions, which are used as an input in the Pb-Pb fit.   
However, because of exceeding computing time needs, it has not been possible to 
assess these errors. 
  
In a simple geometrical picture, the single-interacting nucleons are located 
in the outer corona of the portion of colliding nuclei corresponding to the 
observed number of participants. As the projected (on the collision's transverse 
plane) radial nucleon density is:
\begin{equation} 
 \frac{\d N}{\d r} = 4 \pi r \sqrt{R^2 - r^2} n_0
\end{equation}
where $n_0 = 0.16$ fm$^{-3}$ is the nucleon density and $R \simeq 6.45$ fm is
the radius of the portion of colliding nucleus corresponding to a participant  
number of 180, the 50 single-interacting nucleons should lie between 4.84 and
6.45 fm. This simple estimate is in approximate agreement with the calculations 
performed with the Glauber model \cite{urqmd}.
Since the number of single-interacting nucleons is expected to be weakly 
dependent on center-of-mass energy, the fits to this two-component model 
should yield consistent values of $\langle N_c \rangle$ at 30, 40 and 80$A$ GeV
collisions. However, no measurement of hadron production in NN collisions
at those energies is available and this question cannot be tackled for the 
present.  
  
\subsection{Non-equilibrium of hadrons with light quarks}

An extension of the statistical model has been proposed where QGP hadronization 
is essentially a statistical coalescence of quarks occurring at an energy density 
value which does not correspond to a hadron gas at equilibrium \cite{rafe}.
In this model two non-equilibrium parameters are introduced for the different types 
of quarks, $\gq$ for u, d quarks and $\gss$ for strange quarks (the difference 
between $\gss$ and $\gs$ is explained below). The multiplicity of each 
hadron thus reads:
\begin{equation}\label{meangq}
 \langle n_j \rangle = \frac{(2J_j+1) V}{(2\pi)^3} 
 \sum_{n=1}^\infty \gss^{n n_s} \gq^{n n_q}
 \int \d^3 {\rm p} \; \exp[-n\sqrt{{\rm p}^2 + m_j^2}/T + n \muv \cdot \qj/T]  
\end{equation}
where $n_s$ is the number of valence s quarks and $n_q$ the number of valence u, d 
quarks; $\muv$ and $\qj$ are as in Eq. (\ref{mean}). By defining:
\begin{equation}\label{transform}
     \gs = \frac{\gamma_s}{\gq} \qquad  \tilde V = V \gq^2
\end{equation}
the Boltzmann limit of average multiplicity reads:
\begin{equation}\label{meangq2}
  \langle n_j \rangle = \frac{(2J_j + 1) {\tilde V}}{(2\pi)^3} 
  \gs^{n_s} \gq^{|B_j|} \int \d^3 {\rm p} \; \exp[-\sqrt{{\rm p}^2+m_j^2}/T
  + \muv \cdot \qj/T]
\end{equation}
where $B_j$ is the baryon number, as long as mesons have two and baryons have three 
valence quarks. By comparing this formula with the Boltzmann limit of Eq. (\ref{mean})
it can be realized that the introduction of a light-quark non-equilibrium parameter 
amounts to introduce in the statistical model an overall enhancement (or suppression) 
of baryons with respect to mesons, unlike in the model SHM($\gs$). We henceforth 
refer to this model as SHM($\gs\gq$). 

The parameter 
$\gq$ has a definite physical bound for bosons which can be obtained by requiring 
the convergence of the series $\sum_{N=0}^\infty (\gq^{n_q N}) \exp(-N \epsilon/T 
+ N \muv \cdot \qj/T)$ for any value of the energy. If the number of u, d quarks 
to be hadronized is so large that $\gq$ is to attain its bounding value, a Bose 
condensation of particles in the lowest momentum state sets in. For low 
strangeness and electrical chemical potentials, such as those found in the 
present analysis, the bounding value is $\gq = \exp(m/2T)$ where $m$ is the 
neutral pion mass, e.g. $\gq \simeq 1.5$ for $T \simeq 160$ MeV. 

With the introduction of $\gq$ as an additional free parameter, there are 5 parameters
to be determined in the model. This makes the minimization procedure rather 
unstable because it becomes easier to be trapped in local minima. To avoid this,
we have performed 4 parameter fits with fixed values of $\gq$ varying from 0.6 to
1.7 in steps of 0.1. This method allows a clearcut determination of the absolute
minimum.

\begin{figure}
\caption{Minimum $\chi^2$ of multiplicity fits in Pb-Pb collisions at a beam 
energy of 158$A$ GeV as a function of a fixed light quark non equilibrium 
parameter $\gq$. The round dots are the $\chi^2$'s obtained with the main data 
sample, whilst triangular dots are those obtained with pion multiplicities 
lowered by one standard deviation and all others unchanged. The vertical 
dashed line indicates the condensation point. \label{gammaq}}
\includegraphics[scale=0.9]{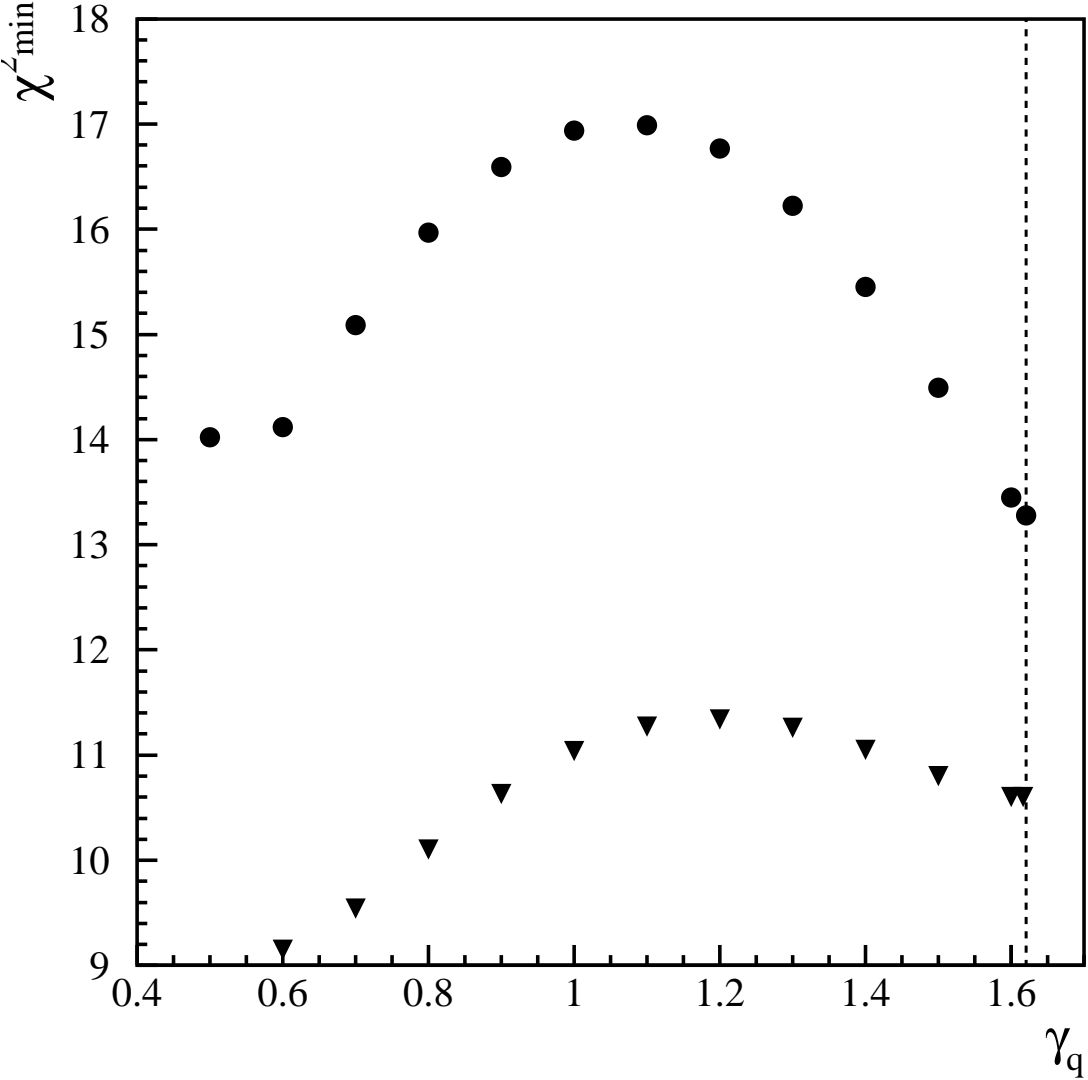}
\end{figure}

The results of these fits are shown in fig.~\ref{gammaq} in terms of the minimum 
$\chi^2$ obtained for fixed $\gq$. The round dots show the minimum $\chi^2$'s achieved 
by using the main sample of multiplicities in Pb-Pb collisions at 158$A$ GeV. 
We find a steady decreasing trend in the value of best-fit temperatures varying 
from $\simeq 187$ MeV at $\gq = 0.6$ to $\simeq 140$ MeV at $\gq=1.6$. The fitted 
temperature at $\gq=0.5$ reaches the upper limit of 200 MeV, which is the maximum 
allowed in the model to prevent from being critically dependent on the hadron mass 
spectrum cut-off. On the other hand, the best-fit values of $\mu_B/T$ and $\gs$ are
rather stable and about the same found in the main fit with $\gq=1$. 
The number of terms in the series (\ref{meangq}) has been truncated to 5 for 
all particles; the contribution of further terms has been found to be negligible 
throughout. 
 
It is seen that the absolute $\chi^2$ minimum falls in the region of pion condensation, 
marked by a vertical line at $\gq \simeq 1.62$, with $\chi^2 \simeq 13$ and $T \simeq
140$ MeV. This finding is in agreement with what is found in ref.~\cite{rafe}. 
However, there is also a local minimum at the lower edge $\gq = 0.6$, with a temperature
of 187 MeV, which is only one unit of $\chi^2$ higher than the absolute minimum. 
This indicates that the absolute minimum could be rather unstable against variations
of the input data and this is in fact what we find by varying down the pion multiplicities
by only 1 $\sigma$. For this case, the minimum $\chi^2$s are shown in fig.~\ref{gammaq} 
as triangular dots and the absolute minimum now lies at $\gq = 0.6$ instead of at
the pion condensation point.  
  
In view of the instability of the fit, and of the small {\em relative} $\chi^2$ 
improvement in comparison with the main fit, we conclude that there is so far 
no evidence for the need of this further non-equilibrium parameter. However, it is
interesting to note that this model predicts an enhanced (if $\gq > 1$) or suppressed
(if $\gq < 1$) production of the $\Theta^+$ pentaquark baryon with respect to the 
other versions of the statistical model, even though they agree in reproducing
the multiplicities of all other hadrons. This is owing to an additional $\gq^2$ 
factor for this special hadron having five valence quarks. From Eqs. 
(\ref{meangq},\ref{transform}) one gets, in the Boltzmann limit:
\begin{equation}\label{theta2}
\langle n_{\Theta^+} \rangle = \frac{\gq^3 \gs \tilde V}{\pi^2} m^2 T 
{\rm K}_2 \left(\frac{m}{T}\right) \exp[\mu_B/T+\mu_Q/T+\mu_S/T]  
\end{equation}
In fact, as can be seen in table~\ref{pbpb158} the predicted yield of $\Theta^+$ 
at the global minimum $\gq \simeq 1.62$ is more than a factor 2 higher than for 
SHM($\gs$). 

As a final remark, we stress that minimum $\chi^2$ fits are very useful tools to get 
information on the state of the source at chemical freeze-out, but, as already
emphasized in Sect.~3, the simple multiplicity analysis with global parameters
resides on an idealization of the collision (e.g. the assumed existence of an EGC) 
which cannot {\em exactly} fit physical reality and discrepancies are to be expected
anyway. Thus, a new mechanism or of a modification of the basic scheme proves to be 
relevant only if it leads to a major improvement of the agreement with the data. 
Slight improvements of the $\chi^2$, whenever their significance is beyond its 
expected statistical fluctuations, cannot be seriously taken as a proof of the 
validity of a particular scheme.  

\section{Energy Dependence} 

The statistical model does not make any prediction on the energy dependence of 
hadron production; its relevant parameters have to be determined separately
for each energy and reaction type. Nevertheless, the analysis of the data within this model
may help in the study of energy dependence of hadron production because it effectively
reduces the full experimental information on numerous hadron yields to only few 
parameters. Clearly this reduction should be taken with care, were not for the
approximate validity of some relevant assumptions, such as the reducibility to
EGC (see Sect.~2). Furthermore, the reduction procedure may remove or dilute 
essential physical 
information. With these caveats in mind, in this section we discuss the energy 
dependence of the chemical freeze-out parameters extracted from the data (by
using the full data sets) within our main SHM($\gs$) approach.

\begin{figure}
\caption{Chemical freeze-out points in the $\mu_B-T$ plane in various heavy ion 
collisions. The full round dots refer to Au-Au at 11.6 and Pb-Pb collisions 
at 40, 80, 158$A$ GeV obtained in the analysis A, whilst the hollow square dot 
has been obtained in ref.~\cite{flork} by using particle ratios measured at 
midrapidity in Au-Au collisions at $\sqrt s_{NN} = 130$ GeV.
The hollow round dot without error bars refers to Pb-Pb collisions at 
30$A$ GeV and has been obtained by forcing $T$ and $\mu_B$ to lie on the
parabola fitted to the full round dots.\label{tmu}}
\includegraphics[scale=0.9]{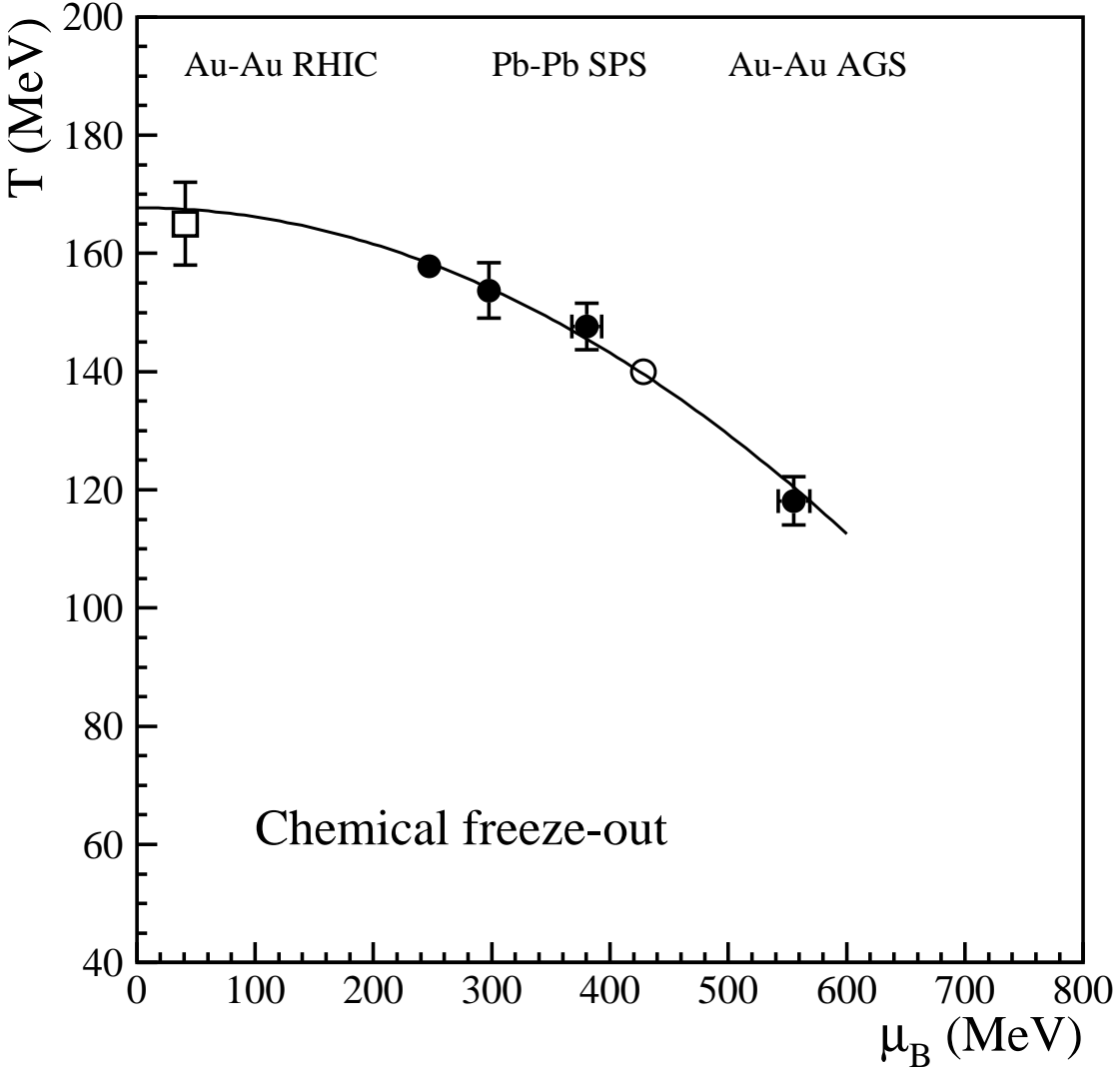}
\end{figure}

The chemical freeze-out points in the $\mu_B-T$ plane are shown in fig.~\ref{tmu}.
The RHIC point at $\sqrt s_{NN} = 130$ GeV, obtained fitting particle yield 
ratios at midrapidity, has been taken from ref.~\cite{flork}. The four points
at beam energies of 11.6, 40, 80 and 158$A$ GeV have been fitted with a parabola:
\begin{equation}
T = 0.167 - 0.153 \mu_B^2,
\label{tmuparam}
\end{equation}
where $T$ and $\mu_B$ are in GeV. The Pb-Pb point at 30$A$ GeV has been forced
to lie on the above curve, as has been mentioned in the previous section. The RHIC
point calculated in ref.~\cite{flork} is in good agreement with the extrapolation
of the curve (\ref{tmuparam}). 

\begin{figure}
\caption{Strangeness non-equilibrium parameter $\gs$ as a function of the nucleon-nucleon
centre-of-mass energy. Full dots refer to fit A, hollow dots to fit B.\label{gs}}
\includegraphics[scale=0.9]{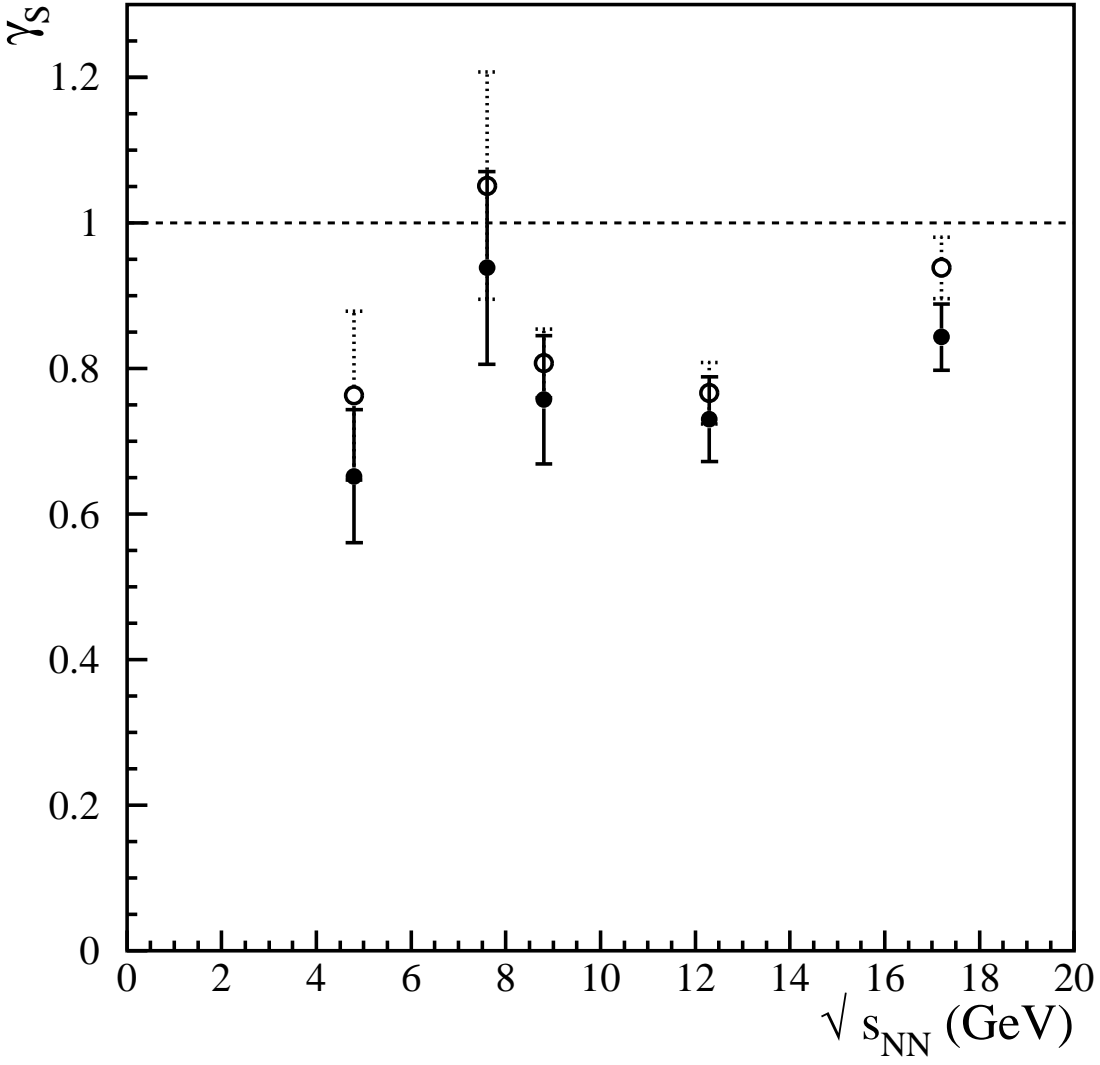}
\end{figure}

In the search for deconfinement phase transition, strangeness production is 
generally believed to be a major item, especially if an anomalous abrupt change
was found as a function of centre-of-mass energy or other related 
quantities. A possible indication of it in Pb-Pb collisions at the low  
SPS energies was reported on the basis of the observed energy dependence of 
several observables \cite{anomalies}. Particularly, the \kpi ratio shows a peaked 
maximum at about 30$A$ GeV. One may expect that this anomaly should be reflected 
in the energy dependence of $\gs$ parameter fitted within SHM($\gs$) scheme.
This dependence is plotted in fig.~\ref{gs} and in fact a maximum shows up at 
30$A$ GeV. Although the error bars are large enough so as to make $\gs$ seemingly 
consistent with a constant as a function of centre-of-mass energy, it is
important to note that the dominant systematic errors on experimental data
at SPS energies are essentially common. Therefore, the errors on the model 
parameters at different SPS energies turn out to be strongly correlated, 
hence fitted $\gs$'s are expected to move up or down together.

\begin{figure}
\caption{Measured \kpi ratio as a function of the fitted baryon-chemical 
potential. The full square dot is a preliminary full phase space measurement
in Au-Au collisions at $\sqrt s_{NN} = 200$ GeV \cite{jordre} and the error
is only statistical; the arrow on the left signifies that its associated 
baryon chemical potential is lower than that estimated at $\sqrt s_{NN} = 130$ 
GeV \cite{flork} used here. For the SPS energy points the statistical 
errors are indicated with solid lines, while the contribution of the common 
systematic error is shown as a dotted line. Also shown the theoretical values 
for a hadron gas along the fitted chemical freeze-out curve shown in fig.~\ref{tmu}, 
for different values of $\gs$. \label{kpi}}
\includegraphics[scale=0.9]{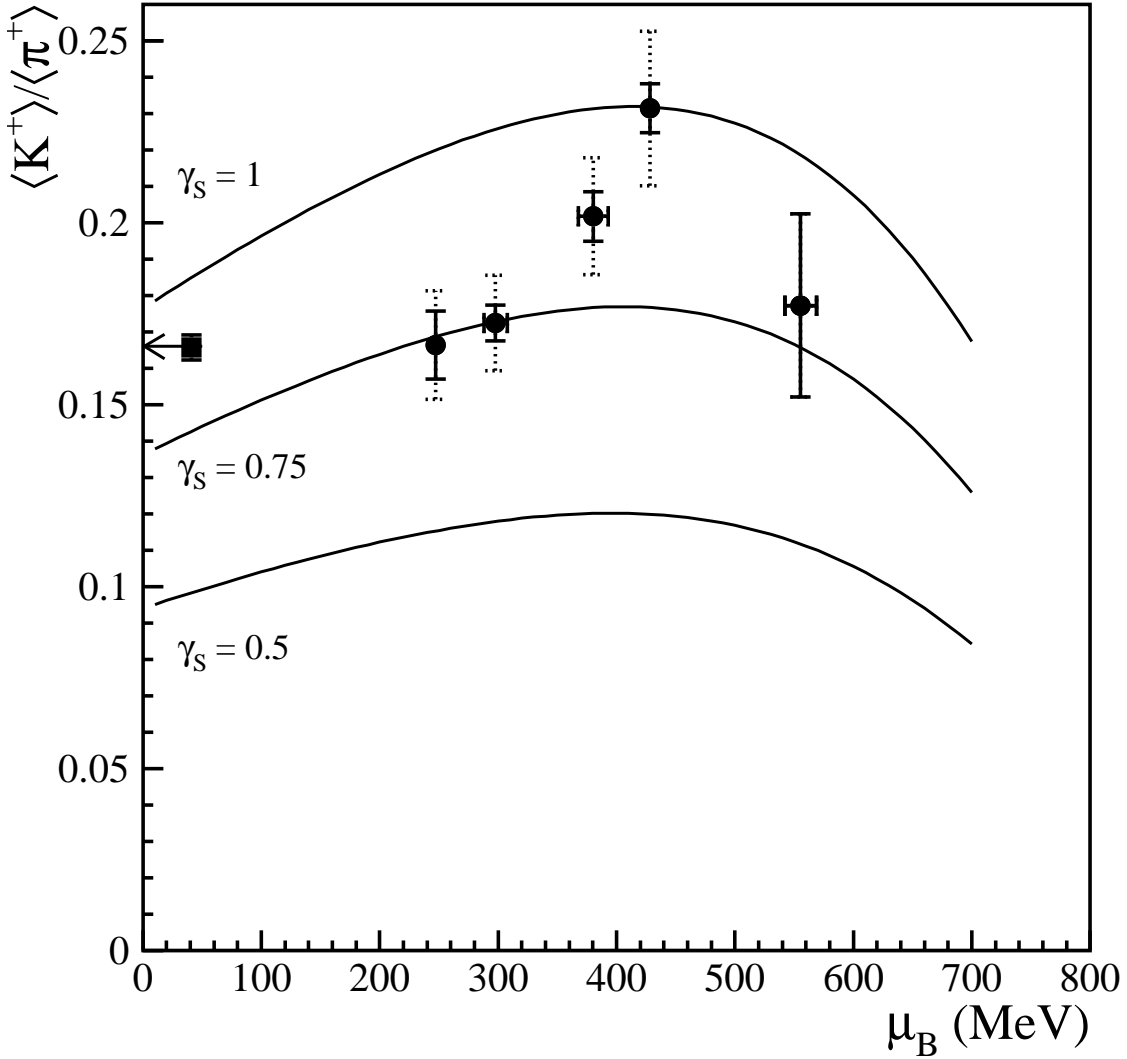}
\end{figure}

In order to further study strangeness production features, we have also compared the
the measured \kpi ratio (including the preliminary RHIC result at $\sqrt s_{NN}
= 200$ GeV \cite{jordre}) with the theoretical values in a hadron gas along 
the freeze-out curve (\ref{tmuparam}) as a function of the fitted baryon-chemical 
potential for different values of $\gs$ (see fig.~\ref{kpi}). The calculated 
dependence of \kpi on 
$\mu_B$ is non-monotonic with a broad maximum at $\mu_B \simeq 400$ MeV (i.e. 
$E_{beam} \simeq 30A$ GeV) \cite{cleymax}. Taking into account that systematic
errors at different energies in Pb-Pb collisions are fully correlated, we can
conclude that the data points seem not to follow the constant $\gs$ lines.

\begin{figure}
\caption{$\ls$ estimated from the fits A (full dots) and B (hollow dots) as a 
function of the fitted baryon-chemical potential. Also shown the theoretical 
values for a hadron gas along the fitted chemical freeze-out curve shown in 
fig.~\ref{tmu}, for different values of $\gs$. \label{ls}}
\includegraphics[scale=0.9]{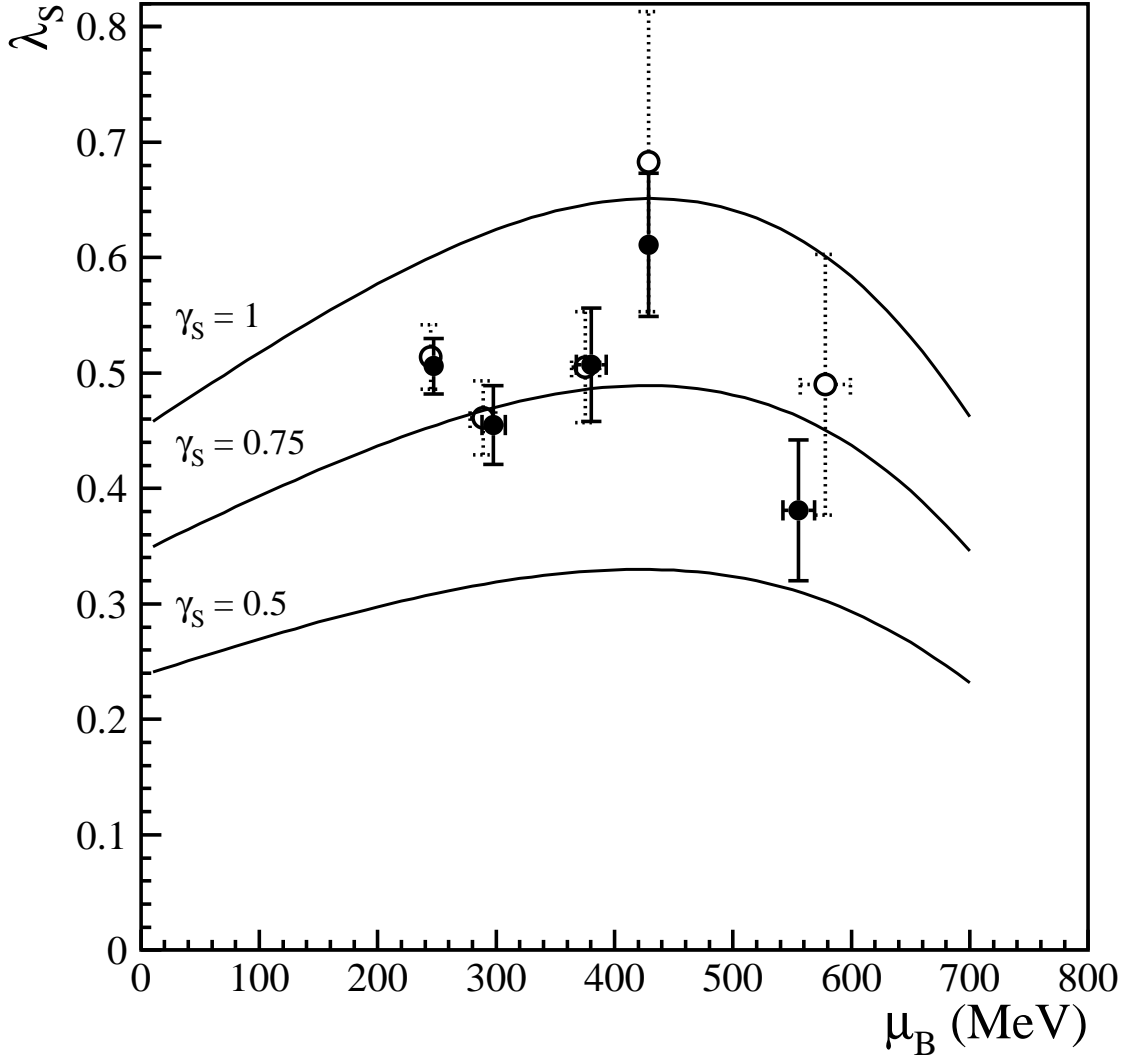}
\end{figure}

In fact, the anomalous increase of relative strangeness production at 30$A$ GeV 
can be seen also in the Wroblewski variable $\ls = 2 \ssb/(\uub+\ddb)$, the estimated 
ratio of newly produced strange quarks to u, d quarks at primary hadron level, 
shown in fig.~\ref{ls} and table~\ref{parameters}. The calculation of newly produced 
quark pairs is performed by using the statistical model best fit values of the 
various hadron multiplicities, so the obtained $\ls$ values are somehow model-dependent. 
Nevertheless, this variable features a very similar behaviour as the ratio \kpi 
and attains a maximum value of 0.61 at 30$A$ GeV, very close to that predicted for 
$\gs=1$.  

These deviations from a smooth behaviour of strangeness production are certainly 
intriguing, yet the analysis within the SHM will be more conclusive in this regard 
with a larger data sample at 30$A$ GeV and at the forthcoming 20$A$ GeV data.         

\section{Summary and Conclusions}

We presented a detailed study of chemical freeze-out conditions in ultra-relativistic
heavy ion collisions at projectile momenta of 11.6 (Au-Au at AGS), 30, 40, 
80 and 158$A$ (Pb-Pb at SPS) GeV, corresponding to nucleon-nucleon centre 
of mass energies of 4.8, 7.6, 8.8, 12.3, 17.2 GeV respectively. By analyzing 
hadronic multiplicities measured in full phase space within the statistical 
hadronization model, we have tested and compared different versions of this model, 
with special emphasis on possible explanations of the observed strangeness 
under-saturation at the hadron level. 

It is found that version of the model referred to as SHM($\gs$), where
a non-equilibrium population of hadron carrying strange valence quarks is allowed,
fits all the data analyzed in this paper. We have also shown that the seeming 
full chemical equilibrium found for central Pb+Pb collisions at 158$A$ GeV by 
using particle yields integrated over a limited region around mid-rapidity is 
most likely an artefact of the kinematical cut.    

We have tested a model (SHM(TC)) in which hadron production is pictured as 
stemming from two independent components: a fireball (or a set of fireballs) at 
full chemical equilibrium and single nucleon-nucleon interactions. This model 
can fit the data at 158$A$ GeV if the number of collisions is around 50 with a
sizeable uncertainty. So far, it cannot be confirmed at other energies due 
to the lack of the precise data on NN interactions.

A model in which strangeness is assumed to vanish locally \cite{redl} yields
a worse fit to the data with respect to SHM($\gs$) and SHM(TC).    

Finally, we have also tested a model in which it is allowed a non-equilibrium
population of hadrons carrying both strange and light valence quarks. We have
found the present set of available data does not allow to establish whether
a further non-equilibrium parameter is indeed needed to account for the 
observed hadron production pattern. A discriminating prediction of this model
with respect to the SHM($\gs$) is an enhanced production of the recently 
discovered $\Theta^+$ pentaquark baryon due to the additional factor $\gamma_q^2$.

Energy dependence of chemical freeze-out parameters has been discussed based
on the results obtained with our main version of the model SHM($\gs$).
The evolution of the freeze-out temperature and baryon-chemical potential
is found to be smooth in the AGS-SPS-RHIC energy range. The strangeness-suppression 
parameter $\gs$ is found to smaller than one ($\gs \simeq 0.8$) for most of 
the studied collisions which confirms previous findings \cite{bgs,beca01,cley}, 
with an indication of a maximum at 30$A$ GeV, where $\gs$ is found to be close 
to one. The significance of this maximum is related to the correlation between
errors on hadron yield measurements at different energies. The interpolated 
dependence of relative strangeness production on energy and baryon-chemical
potential, as measured by the \kpi ratio and the Wroblewski factor $\ls$, 
features a broad maximum at about 30$A$ of beam energy. However, the 
experimental measurement of \kpi and the estimated $\ls$ value in Pb-Pb 
collisions at this energy seemingly exceed the expected values for a fixed
$\gs$.  

\appendix
\section*{APPENDIX A - From canonical to grand-canonical ensemble}

The canonical partition function of the $i^{\rm th}$ cluster can be 
written as a multiple integral over the interval $[-\pi,\pi]$ \cite{beca2}:
\begin{equation}\label{zcan}
 Z_i(\Qi) = \frac{1}{(2\pi)^3} \int \d^3 \phi \; \e^{\i \Qi \cdot \phivs}
 \exp[F(\phiv)]
\end{equation}   
where $\Qi = (Q_i, B_i, S_i)$ is a vector having as components the electric 
charge, baryon number and strangeness of the cluster, $\phiv = 
(\phi_Q,\phi_B,\phi_S)$ and $F(\phiv)$ reads:
\begin{equation}\label{func}
 F(\phiv) = \sum_j \frac{(2J_j+1) V_i}{(2\pi)^3} \int \d^3 {\rm p} \; 
  \log (1\pm \e^{-\sqrt{{\rm p}^2 + m_j^2}/T - \i \phivs \cdot \qj})^{\pm 1}
\end{equation}   
$V_i$ is the volume and $T$ the temperature of the cluster; the sum runs 
over all hadronic species $j$ and $\qj$ the charge vector of the $j^{\rm th}$ 
hadron; the upper sign applies to fermions, the lower to bosons. 
The probability distribution required for the reduction to EGC to apply 
reads \cite{beca2,bgs}:
\begin{equation}\label{weight}
 w(\Q1,\ldots,\QN) = \frac{\prod_i Z_i (\Qi) \delta_{\sum_i \Qi, \Qz}}
 {\sum_{\Q1,\ldots,\QN} \prod_i Z_i (\Qi) \delta_{\sum_i \Qi, \Qz}}
\end{equation}
In this case, the overall multiplicity of the $j^{\rm th}$ hadron is given 
by \cite{beca2}:
\begin{equation}\label{canmean}
 \langle n_j \rangle = \frac{\partial}{\partial \lambda_j} \log Z (\Qz) 
 \Big|_{\lambda_j=1}
\end{equation}
where $Z(\Qz)$ is the canonical partition function of the equivalent global
cluster:
\begin{equation}\label{globz}
  Z(\Qz) = \sum_{\Q1,\ldots,\QN} \prod_i Z_i (\Qi) \delta_{\sum_i \Qi, \Qz}   
\end{equation}
and $\lambda_j$ is a fictitious fugacity parameter.
Formally, this turns out to be the same function as in Eq. (\ref{zcan}) with 
$V=\sum_i V_i$ replacing $V_i$ and $\Qz = \sum_i \Qi$ replacing $\Qi$. 
If $V$ is large, the canonical partition function can be approximated
by the leading term of an asymptotic saddle point expansion. Setting 
$\exp(-\i \phi_k) = z_k$, the canonical partition function can be 
written as:
\begin{equation}
 Z(\Qz) =  \frac{1}{(2\pi\i)^3} \Big[ \prod_{k=1}^3 \oint \frac{\d z_k}{z_k} \Big] 
 \exp[F_c({\bf z})] \prod_{k=1}^3 z_k^{-Q_k}
\end{equation}
where:
\begin{equation}\label{func2}
 F_c({\bf z}) = \sum_j \frac{(2J_j+1) V_i}{(2\pi)^3} \int \d^3 {\rm p} 
 \; \log ( 1\pm \e^{-\sqrt{{\rm p}^2 + m_j^2}/T} \prod_{k=1}^3 z_k^{q_{jk}})^{\pm 1}
\end{equation}   
The saddle-point expansion is carried out by requiring the logarithmic 
derivative of the integrand to vanish:
\begin{equation}\label{saddle}
  - \frac{Q_k}{z_k} + \frac{\partial F_c}{\partial z_k} = 0 \qquad k = 1, 2, 3
\end{equation}
The solutions of this equation are indeed the grand-canonical fugacities 
$\lambda_k \equiv \exp(\mu_k/T)$. The function $F_c(\lambdav)$ coincides 
with the logarithm of the grand-canonical partition function $\log Z_{gc}$, 
therefore the equation (\ref{saddle}) expresses the equality between the 
average charge in the grand-canonical ensemble $\lambda_k \partial 
\log Z_{gc}/\partial \lambda_k$ and the initial value $Q_k$. 
The canonical partition function now becomes, at the second order of the 
expansion:
\begin{equation}\label{cantogc}
 Z(\Qz) \simeq \exp[F_c(\lambdav)] \Big[ \prod_{k=1}^3 \lambda_k^{-Q_k} \Big]
 \frac{1}{(2\pi\i)^3} \Big[ \prod_{k=1}^3 \oint \frac{\d z_k}{z_k} \Big] 
 \exp [-({\bf z}-\lambdav)^T {\sf H} \, ({\bf z}-\lambdav)/2]
\end{equation}
where $\sf H$ is the Hessian matrix in $z_k = \lambda_k$. The first 
exponential factor is just the grand-canonical partition function $Z_{gc}$ 
calculated for the fugacities $\lambda_k$. The average multiplicity of the
$j^{\rm th}$ hadron species can now be calculated by using Eq.~(\ref{canmean})
taking the approximated expression (\ref{cantogc}) of the canonical partition 
function. Retaining only the dominant contribution, one just obtains the 
grand-canonical expression of the average multiplicity as expressed in 
Eq. (\ref{mean}). 

\section*{APPENDIX B - Proof of equation (\ref{scanonical})}

The argument closely follows the previous one. The main difference is the
request of vanishing strangeness for each cluster. Thus, the configurational 
probabilities (\ref{weight}) $w$ turn to:
\begin{equation}
 w(\Q1,\ldots,\QN) = \frac{\prod_i Z_i (\Qi) \delta_{\sum_i \Qi, \Qz} \, 
 \delta_{S_i,0}}
 {\sum_{\Q1,\ldots,\QN} \prod_i Z_i (\Qi) \delta_{\sum_i \Qi, \Qz} \, 
 \delta_{S_i,0}}
\end{equation}
and the average multiplicity of the $j^{\rm th}$ hadron species now reads:
\begin{equation}\label{means}
 \langle n_j \rangle = \frac{\partial}{\partial \lambda_j} 
 \log \sum_{\Q1,\ldots,\QN} \prod_i Z(\Qi) \, \delta_{\Qz,\Sigma_i \Qi}
 \, \delta_{S_i,0} \Big|_{\lambda_j=1}
\end{equation}
where $\lambda_j$ is a fictitious fugacity.
Let us now work out the expression:
\begin{equation}\label{global}
 \zeta = \sum_{\Q1,\ldots,\QN} \prod_{i=1}^N Z_i(\Qi) \, \delta_{\Qz,\Sigma_i \Qi}
 \, \delta_{S_i,0}
\end{equation}
by assuming that all the clusters have the same volume $V_c$ and temperature
$T$. This can be done rewriting the canonical partition functions $Z_i$ like 
in Eq.~(\ref{zcan}), using the integral representation of the Kronecker's delta 
and eliminating the redundant strangeness conservation constraints. Thus,
expanding the vector $\Qz$ in its components, the Eq. (\ref{global}) 
becomes:
\begin{eqnarray}\label{global2}
\zeta = && \sum_{B_1,Q_1,\ldots,B_N,Q_N} \int_{-\pi}^{\pi} \frac{\d \phi_B}{2\pi} 
\int_{-\pi}^{\pi} \frac{\d \phi_Q}{2\pi} \; \e^{\i B \phi_B + \i Q \phi_Q} \,
\e^{-\i \sum_i B_i \phi_B - \i \sum_i Q_i \phi_Q} \nonumber \\ 
&& \times \prod_{i=1}^N \int_{-\pi}^{\pi} \frac{\d \phi_{iB}}{2\pi} \int_{-\pi}^{\pi} 
\frac{\d \phi_{iQ}}{2\pi} \int_{-\pi}^{\pi} \frac{\d \phi_{iS}}{2\pi} \; 
\e^{\i B_i \phi_{iB} + \i Q_i \phi_{iQ}} \exp[F(\phi_{iB},\phi_{iQ},\phi_{iS})]
\end{eqnarray}
where $F$ is the function in Eq. (\ref{func}) with $V_c$ replacing $V_i$.
Note that this function is the same for all clusters, being $T$ and $V_c$ 
constant. We can now carry out the sum over all the integers $B_i, Q_i$ in 
Eq.~(\ref{global2}) and get:
\begin{equation}
 \sum_{B_1,Q_1,\ldots,B_N,Q_N} \e^{-\i \sum_i B_i (\phi_B - \phi_{iB})
 - \i \sum_i Q_i (\phi_Q - \phi_{iQ})} =  
  \prod_{i=1}^N (2 \pi)^3 \, \delta (\phi_B - \phi_{iB}) \, 
  \delta (\phi_Q - \phi_{iQ})
\end{equation}
so that the integration over $\phi_{iB}$ and $\phi_{iQ}$ in Eq.~(\ref{global2}) 
can be easily done and one is left with:
\begin{equation}\label{global3}
\zeta = \int_{-\pi}^{\pi} \frac{\d \phi_B}{2\pi} \int_{-\pi}^{\pi} 
 \frac{\d \phi_Q}{2\pi} \; \e^{\i B \phi_B + \i Q \phi_Q} \prod_{i=1}^N 
 \int_{-\pi}^{\pi} \frac{\d \phi_{iS}}{2\pi} \; \exp[F(\phi_{B},\phi_{Q},\phi_{iS})]
\end{equation}
As the function $F$ is the same for all clusters, this can be written also
as:
\begin{equation}\label{global4}
 \zeta = \int_{-\pi}^{\pi} \frac{\d \phi_B}{2\pi} \int_{-\pi}^{\pi} 
 \frac{\d \phi_Q}{2\pi} \; \e^{\i B \phi_B + \i Q \phi_Q} \left\{ \int_{-\pi}^{\pi} 
 \frac{\d \phi_{S}}{2\pi} \; \exp[F(\phi_{B},\phi_{Q},\phi_{S})] \right\}^{V/V_c}
\end{equation}
being $V = \sum_i V_i = N V_c$. For large volumes, we can approximate $\zeta$ by
means of the saddle-point expansion of the integrals over $\phi_B$ and $\phi_Q$
like in Appendix A. Thus, similarly to Eq. (\ref{cantogc}):
\begin{equation}\label{final}
 \zeta \propto \lambda_B^{-B} \lambda_Q^{-Q} \left\{ \int_{-\pi}^{\pi} 
 \frac{\d \phi_{S}}{2\pi} \; \exp \Big[ \sum_j \frac{(2J_j+1) V_c}{(2\pi)^3} 
 \int \d^3 {\rm p} \; \log (1\pm \lambda_B^{B_j} \lambda_Q^{Q_j} 
 \e^{-\sqrt{{\rm p}^2 + m_j^2}/T -i \phi_S S_j})^{\pm 1} \Big] \right\}^{V/V_c}
\end{equation}
The function:
\begin{equation}\label{zscan}
  Z_c \equiv \int_{-\pi}^{\pi} 
 \frac{\d \phi_{S}}{2\pi} \; \exp \Big[ \sum_j \frac{(2J_j+1) V_c}{(2\pi)^3} 
 \int \d3 {\rm p} \; \log (1\pm \lambda_B^{B_j} \lambda_Q^{Q_j} 
 \e^{-\sqrt{{\rm p}^2 + m_j^2}/T -i \phi_S S_j})^{\pm 1} \Big]
\end{equation}
is defined as the {\em strange canonical partition function}. The multiplicity
of the hadron $j$ can now be calculated by means of Eq. (\ref{means}) by using
Eqs. (\ref{global}),(\ref{global4}) and the definition (\ref{zscan}). What
is obtained is just Eq. (\ref{scanonical})
 
\begin{acknowledgments}
We would like to thank Jens Ivar J\o rdre for his help with \kpi
ratio at RHIC. This work is part of INFN research project FI31.
\end{acknowledgments}



\newpage
\begin{table}
\begin{center}
\caption{Comparison between measured and fitted particle multiplicities, in the framework of
SHM($\gs$) model in central Au-Au collisions (3\%) at a beam energy of 11.6$A$ GeV. Also 
shown the predicted multiplicities of the main hadron species.}\label{ags}
\vspace{0.5cm}
\begin{tabular}{|c|c|c|c|c|}
\hline
                  & Reference            & Measurement          & Fit A   & Fit B         \\
\hline  
 $N_P$            & \cite{centrags}      & $363 \pm  10$	& 361.7   & 360.6    \\
 $p/\pi^+$        & \cite{protags,beca01}& $1.23 \pm 0.13$	& 1.277   & 1.224    \\   
 $\pi^+$          & \cite{pionags,beca01}& $133.7 \pm 9.93$     & 134.9   & 140.0    \\
 $\pi^-$          &                      &                      & 176.9   & 182.2	     \\  
 $\pi^0$          &                  	 &			& 163.5   & 163.2 	     \\
 K$^+$            & \cite{centrags}      & $23.7 \pm 2.86$	& 18.80   & 18.81    \\
 K$^-$            & \cite{centrags}      & $3.76 \pm 0.47$	& 3.890   & 3.539    \\
 K$^0_S$          &                  	 &			& 11.68   & 11.68	     \\
 $\eta$           &                  	 &			& 8.073   & 6.340	     \\
 $\omega$         &                  	 &			& 4.870   & 3.659	     \\
 $\phi$           &                  	 &			& 0.3287  & 0.3489	     \\
 $\eta^{'}$       &                  	 &			& 0.2997  & 0.2437	     \\
 $\rho^+$         &                  	 &			& 7.707   & 10.39	     \\
 $\rho^-$         &                  	 &			& 9.164   & 12.55	     \\
 $\rho^0$         &                  	 &			& 8.517   & 11.59	     \\
 K$^{*+}$         &                  	 &			& 3.555   & 3.512	     \\
 K$^{*-}$         &                  	 &			& 0.6179  & 0.5145	     \\
 K$^{*0}$         &                  	 &			& 3.766   & 3.801	     \\
$\bar{\rm K}^{*0}$&                  	 &			& 0.5555  & 0.4628	     \\
 p                &                  	 &			& 172.2   & 171.4	     \\
 $\bar{\rm p}$    &                  	 &			& 0.02851 & 0.02465	     \\
 $\Delta^{++}$    &                  	 &			& 25.39   & 24.51	     \\
 $\bar\Delta^{--}$&                  	 &			& 0.003071& 0.00222	     \\
 $\Lambda$        & \cite{lamb891,lamb896} see text   & $18.1 \pm 1.9$	 & 19.82    & 20.71    \\
 $\bar\Lambda$    &        	         & $0.017 \pm 0.005$	& 0.01601   & 0.01645	 \\
 $\Sigma^+$       &        	         &			& 4.840     & 4.784	     \\
 $\Sigma^-$       &        	         &			& 5.457     & 5.453	    \\
 $\Sigma0$        &        	         &			& 5.163     & 5.106	    \\
 $\bar\Sigma^-$   &        	         &			& 0.003445  & 0.00321	      \\
 $\bar\Sigma^+$   &        	         &			& 0.002793  & 0.00259	      \\
 $\bar\Sigma^0$   &        	         &			& 0.003115  & 0.00288	      \\ 
 $\Xi^-$          &        	         &			& 0.56067   & 0.5564	     \\
 $\Xi^0$          &        	         &			& 0.54670   & 0.5387	     \\
 $\bar\Xi^+$      &        	         &			& 0.002133  & 0.00248	      \\
 $\bar\Xi^0$      &        	         &			& 0.002392  & 0.00280	      \\
 $\Omega$         &        	         &			& 0.01352   & 0.01459	      \\
 $\bar\Omega$     &        	         &			& 0.0003569 & 0.00056	      \\
 $\Lambda(1520)$  &        	         &			& 0.7720    & 0.6601	     \\
\hline
 $\Theta^+(1540)$     &    	     	 &		        &  1.86        &  2.20	  \\
 $\bar\Theta^-(1540)$ &    	     	 &		        &$2.7 10^{-5}$ &$1.87 10^{-5}$\\
\hline 
\end{tabular}
\end{center}
\end{table}
\newpage
\begin{table}
\begin{center}
\caption{Comparison between measured and fitted particle multiplicities, in the framework of
SHM($\gs$) model, in central Pb-Pb collisions (7.2\%) at a beam energy of 30$A$ GeV. 
Also shown the predicted multiplicities of the main hadron species.}\label{pbpb30}
\vspace{0.5cm}
\begin{tabular}{|c|c|c|c|c|}
\hline
            	   & Reference     & Measurement            & Fit A  &  Fit B \\		 	 
\hline  	  			     						 	 
 $N_P$        	   &\cite{Alt2003rn}  & $349 \pm  1 \pm 5$  &  350.5     & 350.5    \\
 $\pi^+$      	   &\cite{Alt2003rn}  & $239\pm 0.7\pm 17$  &  228.4     & 228.5    \\	
 $\pi^-$      	   &\cite{Alt2003rn}  & $275\pm 0.7\pm 20$  &  256.5     & 256.8    \\	
 $\pi^0$      	   &		      &                     &  265.8     & 251.9    \\	
 K$^+$        	   &\cite{Alt2003rn}  & $55.3\pm 1.6\pm 2.8$&  49.83     & 48.83    \\     
 K$^-$        	   &\cite{Alt2003rn}  & $16.1\pm 0.2\pm 0.8$&  17.11     & 20.72    \\     
 K$^0_S$      	   &		  &     	 	    &  33.57     & 34.85    \\      
 $\eta$       	   &		  &     	   	    &  23.74     & 21.60	   \\	   
 $\omega$     	   &		  &     	   	    &  15.45     & 12.99	   \\	   
 $\phi$            &  	   	  &   		  	    &  2.571     & 2.848	 \\ 
 $\eta^{'}$    	   &  	   	  &   		  	    &  1.411     & 1.341	 \\	
 $\rho^+$     	   &  	   	  &   		  	    &  20.24     & 22.68	 \\	
 $\rho^-$     	   &  	   	  &   		  	    &  23.09     & 26.05	 \\	
 $\rho^0$     	   &  	   	  &   		  	    &  22.14     & 25.11	 \\	
 K$^{*+}$     	   &  	   	  &   		  	    &  13.65     & 13.45	 \\	
 K$^{*-}$     	   &  	   	  &   		  	    &  4.006     & 3.668	 \\	
 K$^{*0}$     	   &  	   	  &   		  	    &  14.21     & 14.20	 \\	
$\bar{\rm K}^{*0}$ &  	   	  &   		  	    &  3.710     & 3.386	 \\    
 p                 &  	   	  &   		  	    &  138.0     & 137.0	 \\    
 $\bar{\rm p}$     &  	   	  &   		  	    &  0.3650    & 0.3803	 \\    
 $\Delta^{++}$     &  	   	  &   		  	    &  26.90     & 25.23	 \\    
 $\bar\Delta^{--}$ &  	   	  &   		  	    &  0.07781   & 0.07438	 \\    
$\Lambda$    	   &  	   	  &   		  	    &  38.02     & 40.25	 \\ 
$\bar\Lambda$	   &  		  &   		  	    &  0.3393    & 0.3901	 \\
 $\Sigma^+$        &		  &     	   	    &  9.995     & 9.935	 \\	 
 $\Sigma^-$        &		  &     	   	    &  10.91     & 10.89	 \\	 
 $\Sigma^0$        &		  &     	   	    &  10.48     & 10.40	 \\	 
 $\bar\Sigma^-$    &		  &     	   	    &  0.1005    & 0.1106	   \\	  
 $\bar\Sigma^+$    &		  &     	   	    &  0.08620   & 0.09491	    \\     
 $\bar\Sigma^0$    &		  &     	   	    &  0.09334   & 0.1023	   \\	  
 $\Xi^-$           &		  &     	   	    &  2.422     & 2.422	 \\	
 $\Xi^0$           &		  &     	   	    &  2.378     & 2.369	 \\	
 $\bar\Xi^+$       &		  &     	   	    &  0.07920   & 0.09332	   \\	  
 $\bar\Xi^0$       &		  &     	   	    &  0.08446   & 0.09967	   \\	  
 $\Omega$          &		  &     	   	    &  0.1587    & 0.1799	  \\	 
 $\bar\Omega$      &		  &     	   	    &  0.02067   & 0.02959	   \\	  
 $\Lambda(1520)$   &		  &     	   	    &  2.167     & 1.751	\\	
\hline
 $\Theta^+(1540)$     &           &               	    &  2.84      &  3.02       \\
 $\bar\Theta^-(1540)$ &           &               	    &  0.0018    &  0.0019     \\
\hline 
\end{tabular}
\end{center}
\end{table}
\newpage
\begin{table}
\begin{center}
\caption{Comparison between measured and fitted particle multiplicities, in the framework of
SHM($\gs$) model, in central Pb-Pb collisions (7.2\%) at a beam energy of 40$A$ GeV. 
Also shown the predicted multiplicities of the main hadron species.}\label{pbpb40}
\vspace{0.5cm}
\begin{tabular}{|c|c|c|c|c|}
\hline
            	   & Reference      & Measurement           & Fit A   & Fit B \\		 	 
\hline  	  			     						 	 
 $N_P$        	   &\cite{Af2002mx} & $349 \pm  1 \pm  5$   & 352.1   & 351.6	\\  
 $\pi^+$      	   &\cite{Af2002mx} & $293 \pm 3 \pm 15$    & 285.5   & 288.3  \\ 		
 $\pi^-$      	   &\cite{Af2002mx} & $322 \pm 3 \pm 16$    & 314.7   & 317.9  \\ 		
 $\pi^0$      	   &		    &			    & 330.4   & 314.9       \\ 		
 K$^+$        	   &\cite{Af2002mx} & $59.1 \pm 1.9 \pm 3$  & 51.22   & 50.61  \\ 		
 K$^-$        	   &\cite{Af2002mx} & $19.2 \pm 0.5 \pm 1.0$& 20.52   & 20.33  \\ 		 
 K$^0_S$      	   &		    &			    & 35.79   & 36.24     \\ 	 	 
 $\eta$       	   &		    &			    & 30.48   & 26.59     \\ 	 	 
 $\omega$     	   &		    &			    & 23.00   & 19.57     \\ 	 	 
 $\phi$            &\cite{Af2002fk} & $2.57 \pm 0.10$	    &  2.641  & 2.644	 \\		   
 $\eta^{'}$    	   &		  &			    &  1.858  & 1.621     \\ 	 	 
 $\rho^+$     	   &		  &			    & 28.84   & 31.11     \\ 	 	 
 $\rho^-$     	   &		  &			    & 32.36   & 35.01     \\ 	 	 
 $\rho^0$     	   &		  &			    & 31.27   & 33.97     \\ 	 	 
 K$^{*+}$     	   &		  &			    & 15.55   & 14.34     \\ 	 	 
 K$^{*-}$     	   &		  &			    &  5.393  & 4.860     \\ 	 	 
 K$^{*0}$     	   &		  &			    & 16.06   & 14.98     \\ 	 	 
$\bar{\rm K}^{*0}$ &		  &			    &  5.045  & 4.535      \\   		 
 p                 &		  &			    &141.7    & 141.9      \\   		 
 $\bar{\rm p}$     &		  &			    &  0.9824 & 0.9784      \\   		 
 $\Delta^{++}$     &		  &			    & 29.05   & 27.03      \\   		 
 $\bar\Delta^{--}$ &		  &			    &  0.2181 & 0.1969      \\   		 
$\Lambda$    	   &\cite{Af2002ub}& $45.6 \pm 3.4$	    & 36.60   & 37.36	 \\		 
$\bar\Lambda$	   &\cite{Af2002ub}& $0.74 \pm 0.06$	    &  0.7223 & 0.7297    \\
 $\Sigma^+$        &		  &			    &  9.655  & 9.221      \\   		 
 $\Sigma^-$        &		  &			    & 10.40   & 9.938      \\   		 
 $\Sigma^0$        &		  &			    & 10.05   & 9.567      \\   		 
 $\bar\Sigma^-$    &		  &			    &  0.2116 & 0.2033       \\  		 
 $\bar\Sigma^+$    &		  &			    &  0.1853 & 0.1783       \\  		 
 $\bar\Sigma^0$    &		  &			    &  0.1985 & 0.1901       \\  		 
 $\Xi^-$           &		  &			    &  2.118   & 1.948      \\  		 
 $\Xi^0$           &		  &			    &  2.089   & 1.917      \\  		 
 $\bar\Xi^+$       &		  &			    &  0.1285  & 0.1207      \\  		 
 $\bar\Xi^0$       &		  &			    &  0.1358  & 0.1277      \\  		 
 $\Omega$          &		  &			    &  0.1364  & 0.1344      \\  		 
 $\bar\Omega$      &		  &			    &  0.02719 & 0.02788      \\  		 
 $\Lambda(1520)$   &		  &			    &  2.273   & 1.688     \\   		 
\hline
 $\Theta^+(1540)$     &           &                         &  2.61    &  2.32       \\
 $\bar\Theta^-(1540)$ &           &                         &  0.0052  &  0.0045     \\
\hline 
\end{tabular}
\end{center}
\end{table}
\newpage
\begin{table}
\begin{center}
\caption{Comparison between measured and fitted particle multiplicities, in the framework of
SHM($\gs$) model, in central Pb-Pb collisions (7.2\%) at a beam energy of 80$A$ GeV. 
Also shown the predicted multiplicities of the main hadron species.}\label{pbpb80}
\vspace{0.5cm}
\begin{tabular}{|c|c|c|c|c|}
\hline
                  & Reference     & Measurement           & Fit A      & Fit B   \\
\hline		   
 $N_P$      	  &\cite{Af2002mx}&  $349 \pm 1 \pm  5$	  & 352.0      & 351.5    \\
 $\pi^+$    	  &\cite{Af2002mx}&  $446 \pm 5 \pm 22$	  & 420.3      & 422.7    \\
 $\pi^-$    	  &\cite{Af2002mx}&  $474 \pm 5 \pm 23$	  & 450.9      & 453.7    \\
 $\pi^0$    	  &               &			  & 485.2      & 457.6	  \\
 K$^+$      	  &\cite{Af2002mx}&  $76.9 \pm 2 \pm 4$	  & 70.72      & 69.87    \\
 K$^-$      	  &\cite{Af2002mx}&  $32.4 \pm 0.6 \pm 1.6$ & 35.96    & 35.96    \\ 
 K$^0_S$    	  &		&			  & 52.80      & 53.77 	  \\
 $\eta$     	  &		&			  & 49.37      & 43.47	  \\
 $\omega$   	  &		&			  & 39.51      & 34.43	 \\
 $\phi$     	  &\cite{Af2002fk} &  $4.37 \pm 0.14$	  & 4.354      & 4.353   \\  
 $\eta^{'}$       &         	&			  & 3.196      & 2.786 	  \\
 $\rho^+$         &         	&			  & 47.42      & 48.92	  \\
 $\rho^-$         &         	&			  & 51.74      & 53.54	  \\
 $\rho^0$         &         	&			  & 50.79      & 52.83	  \\
 K$^{*+}$         &         	&			  & 22.81      & 21.10	  \\
 K$^{*-}$         &         	&			  & 10.43      & 9.600	  \\
 K$^{*0}$         &         	&			  & 23.28      & 21.73	  \\
$\bar{\rm K}^{*0}$&         	&			  & 9.892      & 9.089	  \\
 p                &         	&			  & 141.5      & 142.5	  \\
 $\bar{\rm p}$    &         	&			  & 3.379      & 3.649	  \\
 $\Delta^{++}$    &         	&			  & 30.07      & 28.05	  \\
 $\bar\Delta^{--}$&         	&			  & 0.7623     & 0.7439	  \\
$\Lambda$         &\cite{Af2002ub} &  $47.4 \pm 3.7$      & 42.12      & 42.85    \\
$\bar\Lambda$     &\cite{Af2002ub} &  $2.26 \pm 0.35$     & 2.171      & 2.328   \\
 $\Sigma^+$       &             &  		 	  & 11.23      & 10.67 	 \\
 $\Sigma^-$       &             &  		 	  & 11.82      & 11.23	 \\
 $\Sigma^0$       &             &  		 	  & 11.56      & 10.93	 \\
 $\bar\Sigma^-$   &             &  		 	  & 0.6265     & 0.6348	 \\
 $\bar\Sigma^+$   &             &  		 	  & 0.5643     & 0.5729	 \\
 $\bar\Sigma^0$   &             &  		 	  & 0.5961     & 0.6024	 \\ 
 $\Xi^-$          &             &  		 	  & 2.774      & 2.505	 \\
 $\Xi^0$          &             &  		 	  & 2.758      & 2.485	 \\
 $\bar\Xi^+$      &             &  		 	  & 0.3279     & 0.3154	 \\
 $\bar\Xi^0$      &             &  		 	  & 0.3428     & 0.3299	 \\
 $\Omega$         &             &  		 	  & 0.2154     & 0.2090	 \\
 $\bar\Omega$     &             &  		 	  & 0.06132    & 0.06332 \\
 $\Lambda(1520)$  &             &  		 	  & 2.769      & 2.028	 \\
\hline
 $\Theta^+(1540)$     &         &                         &  2.35      &  2.04   \\
 $\bar\Theta^-(1540)$ &         &                         &  0.022     &  0.021  \\
\hline
\end{tabular}
\end{center}
\end{table}
\newpage 
\begin{table}
\begin{center}
\caption{Comparison between measured and fitted particle multiplicities, in the framework of
various versions of the SHM, in central Pb-Pb collisions (5\%) at a beam energy of 158$A$
GeV. Also shown the predicted multiplicities of the main hadron species. The 
$\Lambda(1520)$ multiplicity was not used in the fits (see text). For the SHM($\gs,\gq$) 
model, the multiplicities are those calculated in the $\gq =1.6$ fit.}\label{pbpb158}
\vspace{0.5cm}
\begin{tabular}{|c|c|c|c|c|c|c|c|}
\hline
   Particle       & Reference     & Measurement & SHM($\gs$) fit A& SHM($\gs$) fit B& SHM(SCV)& SHM(TC)    &  SHM($\gamma_S, \gamma_q$)   \\   	  
\hline	                               								                    
 $N_P$            &\cite{Af2002mx}& $362 \pm 1 \pm  5$    & 363.6     & 363.7     & 362.0   &  364.2     & 362.6  \\	 
 $\pi^+$          &\cite{Af2002mx}& $619 \pm 17 \pm 31$   & 551.5     & 533.2     & 502.7   &  563.4     & 578.7   \\	 
 $\pi^-$          &\cite{Af2002mx}& $639 \pm 17 \pm 31$   & 583.5     & 565.6     & 534.1   &  595.3     & 612.5   \\	 
 $\pi^0$          &               &                       & 638.2     &	576.2     & 585.4   &  661.9     & 661.6   \\	 
 K$^+$            &\cite{Af2002mx}& $103 \pm 5 \pm 5$     & 103.5     & 103.9     & 106.7   &  99.98     & 102.3    \\	 
 K$^-$            &\cite{Af2002mx}& $51.9 \pm 1.9 \pm 3$  & 59.57     & 59.35     & 59.54   &  59.23     & 57.77    \\	
 $K^0_S$          &\cite{Af2002ub}& $81 \pm 4$            & 80.31     & 81.13     & 81.65   &  78.19     & 78.61    \\     
 $\eta$           &               &                       & 70.69     &	62.90     & 67.73   &  76.94     & 62.22    \\	
 $\omega$         &               &                       & 54.93     &	45.34     & 48.75   &  58.75     & 43.01    \\	
 $\phi$           &\cite{Afuu}    & $7.6 \pm 1.1$         &  8.136    & 8.676     & 10.07   &  9.088     & 7.084    \\	
 $\eta^{'}$       &               &        	          &  4.940    &	4.461     & 5.021   &  5.471     & 3.923    \\
 $\rho^+$         &               &        	          & 64.56     & 61.67	  & 57.26   &  67.18     & 54.12    \\
 $\rho^-$         &               &        	          & 69.43     & 66.82	  & 61.93   &  72.29     & 57.04    \\
 $\rho^0$         &  	          &		          & 68.86     & 66.84	  & 61.44   &  73.14     & 57.00    \\
 K$^{*+}$         &  	          &		          & 34.43     & 32.05	  & 35.51   &  32.63     & 27.90    \\
 K$^{*-}$         &  	          &		          & 18.10     & 16.47	  & 17.79   &  17.61     & 14.79    \\
 K$^{*0}$         &  	          &		          & 34.93     & 32.81	  & 36.12   &  33.07     & 27.99    \\
 $\bar{\rm K}^{*0}$& 	          &		          & 17.29     & 15.69	  & 16.96   &  16.84     & 14.17    \\
 p                &               &			  &143.71     & 142.9	  & 138.51  &  147.7     & 144.4    \\
 $\bar{\rm p}$    &               &			  &  7.053    & 6.877	  & 5.756   &  7.721     & 7.046  \\
 $\Delta^{++}$    &               &			  & 31.01     & 28.32	  & 29.70   &  30.70     & 29.27  \\
 $\bar\Delta^{--}$&               &			  &  1.595    & 1.393	  & 1.295   &  1.716     & 1.472  \\
 $\Lambda$        &\cite{Af2002ub}& $53.0 \pm 5.0$        & 53.88     & 56.22     & 57.06   &  49.53     & 53.38  \\
 $\bar\Lambda$    &\cite{Af2002ub}& $4.64 \pm 0.32$       &  4.976    & 5.077     & 4.698   &  4.899     &  4.878 \\  
 $\Sigma^+$       &               &			  & 14.45     &	14.11     & 15.28   &  13.21     &  14.57 \\
 $\Sigma^-$       &               &			  & 15.04     &	14.71     & 15.98   &  13.73     &  14.80 \\
 $\Sigma^0$       &               &			  & 14.78     &	14.39     & 15.67   &  13.57     &  14.73 \\
 $\bar\Sigma^-$   &               &			  &  1.424    &	1.375     & 1.348   &  1.392     &  1.396 \\
 $\bar\Sigma^+$   &               &			  &  1.301    &	1.256     & 1.224   &  1.277     &  1.288 \\
 $\bar\Sigma^0$   &               &			  &  1.364    &	1.313     & 1.288   &  1.344     &  1.345 \\
 $\Xi^-$          &\cite{Af2002he}& $4.45 \pm 0.22$       &  4.4581   & 4.335     & 4.757   &  4.446     & 4.650  \\    
 $\Xi^0$          &               &         	          &  4.446    &	4.315     & 4.736   &  4.440     & 4.681  \\    
 $\bar\Xi^+$      &\cite{Af2002he}& $0.83 \pm 0.04$       &  0.8159   & 0.7931    & 0.8234  &  0.8186    & 0.8263  \\    
 $\bar\Xi^0$      &               &   	     		  &  0.8485   &	0.8264    & 0.8578  &	 0.8508  & 0.8593 \\    
 $\Omega$         &\cite{Af2002fk}& $0.62 \pm 0.09$       &  0.4499   & 0.4906    & 0.4414  &	 0.5165  & 0.4299 \\    
 $\bar\Omega$     &\cite{Af2002fk}& $0.20 \pm 0.03$       &  0.1702   & 0.1884    & 0.1690  &	 0.1859  & 0.1535 \\    
 $\Lambda(1520)$  &\cite{Friese}  & $1.57 \pm 0.44$       &  3.669    & 2.724     & 3.889   &	 3.382   & 3.079  \\    
\hline
 $\Theta^+(1540)$     &           &                       &  2.68     & 2.41      &         &	         & 5.71  \\
 $\bar\Theta^-(1540)$ &           &                       &  0.061    & 0.053     &         &	         & 0.13  \\
\hline  
\end{tabular}
\end{center}
\end{table}
\newpage
\begin{table}
\caption{Integrated multiplicities over limited rapidity windows around midrapidity
obtained by fitting the measured distributions with single (G) or double (G+G) gaussians
The yields at 10\% and 20\% centrality have been then multiplied by 1.08 and 1.32 
respectively to convert them at 5\% centrality trigger condition..}\label{yfit}
\vspace{0.5cm}
\begin{tabular}{|c|c|c|c|c|c|c|}
\hline
 Particle      &Centrality& Reference &Fitting function & $\Delta y = 2$  & $\Delta y = 1$ &$\chi^2$/dof  \\
\hline
 $\pi^-$       &   5\%    &  \cite{Af2002mx} & G+G  &    333.16       & 176.82 	& 9.37      \\
 K$^+$         &   5\%    & \cite{Af2002mx}  & G+G  &    57.16        & 29.81  	& 3.33      \\  
 K$^-$         &   5\%    & \cite{Af2002mx}  & G+G  &    32.24        & 16.90  	& 1.27      \\
 $\phi$        &   5\%    & \cite{Afuu}      & G    &    4.327        & 2.35    & 0.04      \\
 $\Lambda$     &   10\%   & \cite{Af2002ub}  & G    &    22.56        & 11.52  	& 0.49      \\
 $\bar\Lambda$ &   10\%   &  \cite{Af2002ub} & G    &    3.039        & 1.723   & 0.58      \\
 $\Xi^-$       &   10\%   &  \cite{Af2002he} & G    &    2.75	      & 1.484   & 1.2	    \\  
 $\Xi^+$       &   10\%   &  \cite{Af2002he} & G    &    0.571        & 0.3314  & 0.71      \\  
 $\Omega$      &   20\%   &  \cite{Af2002fk} & G    &    0.3095       & 0.173   & 0.73      \\  
 $\bar\Omega$  &   20\%   &  \cite{Af2002fk} & G    &    0.126        & 0.0789  & 2.70      \\ 
\hline
\end{tabular}
\end{table}
\newpage
\begin{table}
\caption{Summary of fitted parameters in various heavy ion collisions at AGS and 
SPS in the framework of the SHM($\gs$) model. The 'common set' parameters have been 
obtained by fitting to the measured multiplicities of $\pi^+$, K$^+$, K$^-$, $\Lambda$, 
$\bar\Lambda$ and the participant nucleons in each collision. Also quoted minimum 
$\chi^2$'s, the estimated radius of the EGC and the $\ls$ parameter (see Sect.~4). 
The re-scaled errors (see text) are quoted within brackets. For Pb-Pb at 30 $A$ GeV 
of beam energy, we have constrained $T$ and $\mu_B$ to lie on the fitted chemical 
freeze-out curve, as described in Sect. 4.}\label{parameters}
\vspace{0.5cm}
\begin{tabular}{|c|c|c|c|c|}
\hline
 Parameters     & Main analysis A         & Main analysis B          &  Common set A            & Common set B  \\
\hline
\multicolumn{5}{|c|}{Au-Au 11.6$A$ GeV} \\
\hline               
$T$ (MeV)       & 118.1$\pm$3.5 (4.1)     & 119.1$\pm$4.0 (5.4)      &  119.2$\pm$2.1 (2.9)     & 119.1$\pm$4.0 (6.6)    \\
$\mu_B$ (MeV)   & 555$\pm$12 (13)         & 578$\pm$15 (21)          &  556$\pm$12 (17)         & 576.9$\pm$17.2 (29)    \\
$\gs$           & 0.652$\pm$0.069 (0.079) & 0.763$\pm$0.086 (0.12)   &  0.645$\pm$0.042 (0.058) & 0.761$\pm$0.090 (0.15)\\
\vexp           & 1.94$\pm$0.21 (0.24)    & 1.487$\pm$0.18 (0.25)    &  1.97$\pm$0.12 (0.17)    & 1.494$\pm$0.21 (0.35)    \\
\hline
$\chi^2$/dof    & 4.0/3                   & 5.5/3                    &  3.86/2                  & 5.5/2 \\
$R$ (fm)        & 9.31$\pm$0.69 (0.80)    & 8.32$\pm$0.72 (0.97)     &                          &       \\
$\ls$           & 0.381$\pm$0.053 (0.061) & 0.490$\pm$0.084 (0.11)   & 0.401$\pm$0.053 (0.074)  & 0.487$\pm$0.089 (0.15)   \\
\hline
\multicolumn{5}{|c|}{Pb-Pb 30$A$ GeV} \\
\hline               
$T$ (MeV)       & 139.5                   & 140.3 & & \\
$\mu_B$ (MeV)   & 428.6                   & 428.7 & & \\ 
$\gs$           & 0.938$\pm$0.078 (0.13)  & 1.051$\pm$0.103 (0.16) & & \\
\vexp           & 6.03$\pm$0.50 (0.85)    & 5.273$\pm$0.526 (0.80) & & \\ 
\hline
$\chi^2$/dof    & 5.75/2                  & 4.6/2 & & \\
$\ls$           & 0.611$\pm$0.037 (0.062) & 0.683$\pm$0.086 (0.13) & & \\ 
\hline
\multicolumn{5}{|c|}{Pb-Pb 40$A$ GeV} \\
\hline               
$T$ (MeV)       & 147.6$\pm$2.1 (4.0)     & 145.5$\pm$1.9 (3.5)      &  148.6$\pm$2.1 (4.7)      &  146.1$\pm$2.0 (4.0)  \\
$\mu_B$ (MeV)   & 380.3$\pm$6.5 (13)      & 375.4$\pm$6.4 (12)       &  393$\pm$10 (22)          &  390$\pm$10 (21) \\
$\gs$           & 0.757$\pm$0.024 (0.046) & 0.807$\pm$0.025 (0.047)  &  0.874$\pm$0.064 (0.14)   &  0.961$\pm$0.079 (0.16)   \\
\vexp           & 8.99$\pm$0.37 (0.71)    & 8.02$\pm$0.34 (0.63)     &  8.09$\pm$0.55 (1.24)     &  7.08$\pm$0.53 (1.1)  \\
\hline
$\chi^2$/dof    & 14.7/4                   & 13.6/4                  & 10.1/2                    &  8.1/2 \\
$R$ (fm)        & 8.37$\pm$0.32 (0.61)     & 8.37$\pm$0.31 (0.58)    &                           &       \\
$\ls$           & 0.507$\pm$0.025 (0.049)  & 0.505$\pm$0.026 (0.048) & 0.621$\pm$0.064 (0.14)    &  0.626$\pm$0.071 (0.14) \\
\hline
\multicolumn{5}{|c|}{Pb-Pb 80$A$ GeV} \\
\hline                
$T$ (MeV)       & 153.7$\pm$2.8 (4.7)      & 151.9$\pm$3.4 (5.4)     &  154.6$\pm$3.3 (7.2)     & 152.2$\pm$3.5 (7.5)     \\
$\mu_B$ (MeV)   & 297.7$\pm$5.9 (9.8)      & 288.9$\pm$6.8 (11)      &  300.7$\pm$9.4 (21)      & 292.8$\pm$9.0 (19)    \\
$\gs$           & 0.730$\pm$0.021 (0.035)  & 0.766$\pm$0.026 (0.042) &  0.741$\pm$0.057 (0.13)  & 0.782$\pm$0.061 (0.13)  \\
\vexp           & 15.38$\pm$0.61 (1.0)     & 14.12$\pm$0.65 (1.1)    &  15.0$\pm$1.0 (2.3)      & 13.7$\pm$0.95 (2.0)   \\
\hline
$\chi^2$/dof    & 11.0/4                   & 10.4/4                  & 9.6/2                    & 9.3/2 \\
$R$ (fm)        & 9.03$\pm$0.41 (0.68)     & 9.05$\pm$0.44 (0.71)    &                          &       \\
$\ls$           & 0.455$\pm$0.020 (0.034)  & 0.461$\pm$0.020 (0.032) & 0.482$\pm$0.053 (0.12)   & 0.4568$\pm$0.044 (0.095) \\
\hline
\multicolumn{5}{|c|}{Pb-Pb 158$A$ GeV} \\
\hline                
$T$ (MeV)       & 157.8$\pm$1.4 (1.9)      & 154.8$\pm$1.4 (2.1)      & 156.6$\pm$2.3 (3.3)     & 152.7$\pm$2.1 (3.2) \\
$\mu_B$ (MeV)   & 247.3$\pm$5.2 (7.2)      & 244.5$\pm$5.0 (7.8)      & 238.6$\pm$7.1 (10)      & 232.4$\pm$7.7 (12)      \\
$\gs$           & 0.843$\pm$0.024 (0.033)  & 0.938$\pm$0.027 (0.042)  & 0.722$\pm$0.053 (0.077) & 0.764$\pm$0.065 (0.097)   \\
\vexp           & 21.13$\pm$0.80 (1.1)     & 18.46$\pm$0.69 (1.1)     & 23.2$\pm$1.4 (2.0)      & 21.1$\pm$1.4 (2.2)   \\
\hline
$\chi^2$/dof    & 16.9/9                   & 21.6/9                   & 4.2/2                   & 4.5/2   \\
$R$ (fm)        & 9.41$\pm$0.26 (0.35)     & 9.44$\pm$0.25 (0.39)     &                         &         \\
$\ls$           & 0.506$\pm$0.018 (0.024)  & 0.514$\pm$0.018 (0.028)  & 0.426$\pm$0.037 (0.054) & 0.401$\pm$0.039 (0.058)   \\
\hline               
\end{tabular}
\end{table}
\newpage
\begin{table}
\caption{Fit results in Pb-Pb at 158$A$ GeV with particle yields in 
limited rapidity window.}\label{deltay2}
\vspace{0.5cm}
\begin{tabular}{|c|c|c|c|c|}
\hline
  Particle             &  WA97 measured  &  WA97 fitted    &  NA49 measured    &  NA49 fitted \\
\hline					 
 h$^-$                 &   178$\pm$22    &   157.2         &                   &                \\
 $\pi^-$               &                 &                 &  176.8$\pm$9.8    &  151.8   \\
 K$^+$                 &                 &                 &  29.81$\pm$2.05   &  30.63   \\
 K$^-$                 &                 &		   &  16.90$\pm$1.16   &  18.24   \\
 K$^0_S$               &   21.9$\pm$2.4  &   22.97         &                   &                \\
 $\phi$                &                 &                 &  2.35$\pm$0.34    &  2.900   \\
 $\Lambda$             &   13.7$\pm$0.9  &   13.75         &  12.44$\pm$1.17   &  15.35   \\
 $\bar\Lambda$         &   1.8$\pm$0.2   &   1.837         &  1.86$\pm$0.13    &  1.905   \\
 $\Xi^-$  	       &   1.5$\pm$0.1   &   1.525  	   &  1.603$\pm$0.079  &  1.502    \\
 $\bar\Xi^+$           &   0.37$\pm$0.06 &   0.3782	   &  0.358$\pm$0.017  &  0.3585  \\
 $\Omega$              &                 &		   &  0.228$\pm$0.033  &  0.1858   \\
 $\bar\Omega$	       &                 &		   &  0.104$\pm$0.016  &  0.08894   \\
 $\Omega$+$\bar\Omega$ &   0.41$\pm$0.08 &   0.3136        &                   &           \\
\hline               
\end{tabular}
\end{table}
\begin{table}
\caption{Fit results in Pb-Pb at 158$A$ GeV with particle yields in 
limited rapidity window.}\label{deltay}
\vspace{0.5cm}
\begin{tabular}{|c|c|c|c|}
\hline
  Parameters    & $\Delta y = 1$         &  $\Delta y = 2$         &  WA97 central \\
\hline 
$T$ (MeV)       & 162.7$\pm$2.7 (5.1)    &  161.0$\pm$2.6 (4.2)    &  161.3$\pm$5.4    \\
$\mu_B$ (MeV)   & 229$\pm$12 (23)        &  223$\pm$13 (21)        &  218$\pm$19       \\
$\gs$           & 0.971$\pm$0.044 (0.083)&  0.950$\pm$0.043 (0.070)&  1.085$\pm$0.079  \\
\vexp           & 5.55$\pm$0.31 (0.58)   &  10.71$\pm$0.59 (0.96)  &  4.73$\pm$0.52    \\
\hline
$\chi^2$/dof    &  21.1/6                &  16.0/6                 &  2.7/3            \\
\hline               
\end{tabular}
\end{table}
\begin{table}
\caption{Fit results in Pb-Pb at 158$A$ GeV with different models, as
described in the text: SHM(SCV) (Strangeness Correlation Volume), SHM(TC) 
(Two Component model), SHM($\gamma_S, \gamma_q$) (light quark non-equilibrium model). 
Free fit parameters are quoted along with resulting minimum $\chi^2$'s. The re-scaled 
errors (see text) are quoted withing brackets. For the SHM($\gamma_S, \gamma_q$), 
the fit has been done by fixing $\gq = 1.6$ near the absolute $\chi^2$ minimum 
(see Section 3.2).}\label{models}
\vspace{0.5cm}
\begin{tabular}{|c|c|c|c|}
\hline
  Parameters          &      SHM(SCV)	          &  SHM(TC)	          & SHM($\gamma_S, \gamma_q$) 	
   \\
\hline
$T$ (MeV)             & 157.9$\pm$1.6 (3.3)       & 154.8$\pm$1.5 (1.9)   & 140.4$\pm$1.1 (1.3)   \\
$\mu_B$ (MeV)         & 261.5$\pm$2.4 (4.9)       & 237.1$\pm$7.0 (8.6)   & 218.1$\pm$4.3 (5.2)  \\
$\gs$                 & 1.0 (fixed)	          & 1.0 (fixed)           & 0.929$\pm$0.027 (0.033) \\
\vexp                 & 18.62$\pm$0.52 (1.1)      & 15.50$\pm$0.54 (0.67) & 16.82$\pm$0.59 (0.72) \\
$f$                   & 0.0253$\pm$ 0.0067 (0.014)&		          &		   \\
$\gq$                 &		                  &		          & 1.6 (fixed)    \\
$\langle N_c \rangle$ &		                  & 52.0$\pm$7.8 (9.6)    &  	      \\
\hline
$\chi^2$/dof          & 37.2/9	                  & 13.7/9  	          & 13.4/9	      \\
\hline               
\end{tabular}
\end{table}
\newpage
\begin{table}
\caption{Fitted parameters and multiplicities in pp collisions at a beam energy of 158 GeV,
corresponding to $\sqrt s = 17.2$ GeV. The re-scaled errors (see text) are quoted withing 
brackets.}\label{pp160}
\vspace{0.5cm}
\begin{tabular}{|c|c|c|c|}
\hline
  Parameter       & \multicolumn{3}{|c|}{Value}             \\
\hline
$T$ (MeV)         &\multicolumn{3}{|c|}{187.2$\pm$6.1 (9.3)}     \\
$VT^3$            &\multicolumn{3}{|c|}{5.79$\pm$0.85 (1.3)}     \\
$\ssb$            &\multicolumn{3}{|c|}{0.381$\pm$0.021 (0.032)} \\
\hline
$\chi^2$/dof      &\multicolumn{3}{|c|}{16.1/7}                 \\
$\ls$             &\multicolumn{3}{|c|}{0.224$\pm$0.019 (0.024)}  \\
\hline
  Particle        &  Reference         & Measurement        & Fit             \\
\hline  
 $\pi^+$          & \cite{Bachler:hu}  &  3.15$\pm$0.16     &  3.257    \\
 $\pi^-$          & \cite{Bachler:hu}  &  2.45$\pm$0.12     &  2.441    \\
 $\pi^0$          &	               &                    &  3.317    \\
 K$^+$            & \cite{Bachler:hu}  &  0.21$\pm$0.02	    &  0.1901   \\
 K$^-$            & \cite{Bachler:hu}  &  0.13$\pm$0.013    &  0.09981  \\
 $K^0_S$          & \cite{Bachler:hu}  &  0.18$\pm$0.04	    &  0.1382   \\
 $\eta$           &	               &                    &  0.3918   \\
 $\omega$         &	               &                    &  0.3514   \\
 $\phi$           & \cite{Afuu}        &  0.012$\pm$0.0015  &  0.01593  \\
 $\eta^{'}$       &	               &                    &  0.02576  \\
 $\rho^+$         &	               &                    &  0.4736   \\
 $\rho^-$         &	               &		    &  0.3118   \\
 $\rho^0$         &	               &		    &  0.4254   \\
 K$^{*+}$         &	               &		    &  0.07360   \\
 K$^{*-}$         &	               &		    &  0.02976   \\
 K$^{*0}$         &	               &		    &  0.06192   \\
$\bar{\rm K}^{*0}$&	               &		    &  0.03383   \\
 p                &	               &     	            &  1.126     \\	   
 $\bar{\rm p}$    & \cite{Bachler:hu}  &0.040$\pm$0.007	    &  0.04364   \\
 $\Delta^{++}$    &	               &		    &  0.2937    \\
 $\bar\Delta^{--}$&	               &		    &  0.007650  \\
 $\Lambda$        & \cite{Billmeier}   &0.115$\pm$0.012     &  0.1123    \\
 $\bar\Lambda$    & \cite{Billmeier}   &0.0148$\pm$0.0019   &  0.01453   \\
 $\Sigma^+$       &       	       &		    &  0.03480   \\
 $\Sigma^-$       &	               &		    &  0.02310   \\
 $\Sigma^0$       &	               &		    &  0.03004   \\
 $\bar\Sigma^-$   &	               &		    &  0.003317  \\
 $\bar\Sigma^+$   &	               &		    &  0.004384  \\
 $\bar\Sigma^0$   &	               &		    &  0.003989  \\
 $\Xi^-$          &                    &                    &  0.001874   \\
 $\Xi^0$          &                    &                    &  0.002119   \\
 $\bar\Xi^+$      &                    &                    &  0.0006902 \\
 $\bar\Xi^0$      &	               &                    &  0.0006376 \\
 $\Omega$         &                    &                    &  0.00003783 \\
 $\bar\Omega$     &                    &                    &  0.00002908 \\
 $\Lambda(1520)$  & \cite{Afqj}        & 0.012$\pm$0.003    &  0.009155   \\
\hline
 $\Theta^+$       &                    &                    &  0.005224  \\
 $\bar\Theta^-$   &                    &                    &  0.0001515  \\
\hline
\end{tabular}
\end{table}
\clearpage

\newpage

\newpage

\newpage

\newpage

\newpage

\newpage

\newpage

\newpage

\newpage

\newpage

\newpage

\newpage

\newpage

\end{document}